 \documentclass[final,3p,11pt]{elsarticle}

\makeatletter
\def\ps@pprintTitle{%
 \let\@oddhead\@empty
 \let\@evenhead\@empty
 \def\@oddfoot{\centerline{\thepage}}%
 \let\@evenfoot\@oddfoot}
\makeatother
\usepackage{color}
\usepackage[usenames,dvipsnames]{xcolor}
\usepackage{hyperref}

\biboptions{numbers,sort&compress}
\usepackage{graphicx}
\usepackage{natbib}
\usepackage{amsfonts,amssymb,amsbsy,amsmath,mathtools}

\begin{document}
\begin{frontmatter}

\renewcommand{\thefootnote}{\fnsymbol{footnote}}
\title{\textbf{Two-scale constitutive modeling of a lattice core sandwich beam\let\thefootnote\relax\footnote{{\color{Blue}\textbf{Recompiled, unedited accepted manuscript}}. \copyright 2018. Made available under \href{https://creativecommons.org/licenses/by-nc-nd/4.0/}{{\color{Blue}\textbf{\underline{CC-BY-NC-ND 4.0}}}}}}}

\author[add1,add2]{Anssi T. Karttunen\corref{cor1}}
\cortext[cor1]{Corresponding author. anssi.karttunen@iki.fi. \textbf{Cite as}: \textit{Compos. Part B-Eng} 2019;160:66--75 \href{https://doi.org/10.1016/j.compositesb.2018.09.098}{{\color{OliveGreen}\textbf{\underline{doi link}}}}}
\author[add2]{J.N. Reddy}
\author[add1]{Jani Romanoff}

\address[add1]{Aalto University, Department of Mechanical Engineering, FI-00076 Aalto, Finland}
\address[add2]{Texas A\&M University, Department of Mechanical Engineering, College Station, TX 77843-3123, USA}

\begin{abstract}
Constitutive equations are derived for a 1-D micropolar Timoshenko beam made of a web-core lattice material. First, a web-core unit cell is modeled by discrete classical constituents, i.e., the Euler--Bernoulli beam finite elements (FE). A discrete-to-continuum transformation is applied to the microscale unit cell and its strain energy density is expressed in terms of the macroscale 1-D beam kinematics. Then the constitutive equations for the micropolar web-core beam are derived assuming strain energy equivalence between the microscale unit cell and the macroscale beam. A micropolar beam FE model for static and dynamic problems is developed using a general solution of the beam equilibrium equations. A localization method for the calculation of periodic classical beam responses from micropolar results is given. The 1-D beam model is used in linear bending and vibration problems of 2-D web-core sandwich panels that have flexible joints. Localized 1-D results are shown to be in good agreement with experimental and 2-D FE beam frame results.
\end{abstract}

\end{frontmatter}


\section{Introduction}
Advances in laser welding have brought lightweight all-steel sandwich panels to the market in the past two decades. The panels consist of a structural core (e.g. I- or X-core) between two faces; the joining is done by welds that penetrate through the faces into the core. Laser-welded sandwich panels are currently used, for example, in shipbuilding, but only as minor parts such as staircase landings and non-structural walls \cite{roland1997,kujala2005}. However, increased knowledge of the limit state behavior of the panels is making way for more demanding applications like ship decks \cite{kolsters2010,jelovica2012,jelovica2013,jiang2014,jelovica2014,frank2013,remes2017,gallo2018}. By one estimate \cite{kujala2005}, a ship deck constructed of steel sandwich panels offers 30--50\% weight savings compared to traditional stiffened steel plate solutions. While this study is motivated by ship structures, steel sandwich panels show good potential for applications in bridges and buildings as well \cite{bright2004,bright2007,nilsson2017,briscoe2011}.

To analyze the global structural response of a large ship within computational limits, any deck constructed of sandwich panels needs to be modeled in a homogenized sense without accounting for every small detail. To this end, a sandwich panel may be modeled as an equivalent single-layer (ESL) beam or plate based on the first-order shear deformation theory (FSDT) \cite{reddy2004}. In this study, the focus is on web-core (I-core) sandwich panels for which different ESL-FSDT models based on classical \cite{romanoff2007a,romanoff2007b}, couple-stress \cite{romanoff2014,romanoff2016,gesualdo2017,penta2017} and micropolar \cite{karttunen2018a} continuum theories have been developed. Recently, analytical solutions founded on discrete classical models have also been formulated for web-core plates \cite{pydah2016,pydah2017,pydah2018}. As for the different ESL-FSDT models, it has been shown that only the micropolar approach \cite{karttunen2018a} can correctly capture the deformations of a rigid-jointed web-core structure because it considers both symmetric and antisymmetric shear deformations, as explained schematically in Fig.~1. In this paper, we develop the micropolar approach further by deriving the constitutive relations for a web-core beam with flexible joints via a two-scale energy method. In Fig.~1, the beam and unit cell lengths $L$ and $l$ represent the \textit{macroscale} and \textit{microscale}, respectively.
\begin{figure}[h!]
\centering
\includegraphics[trim={0 0.1cm 0 0.38cm},scale=0.96]{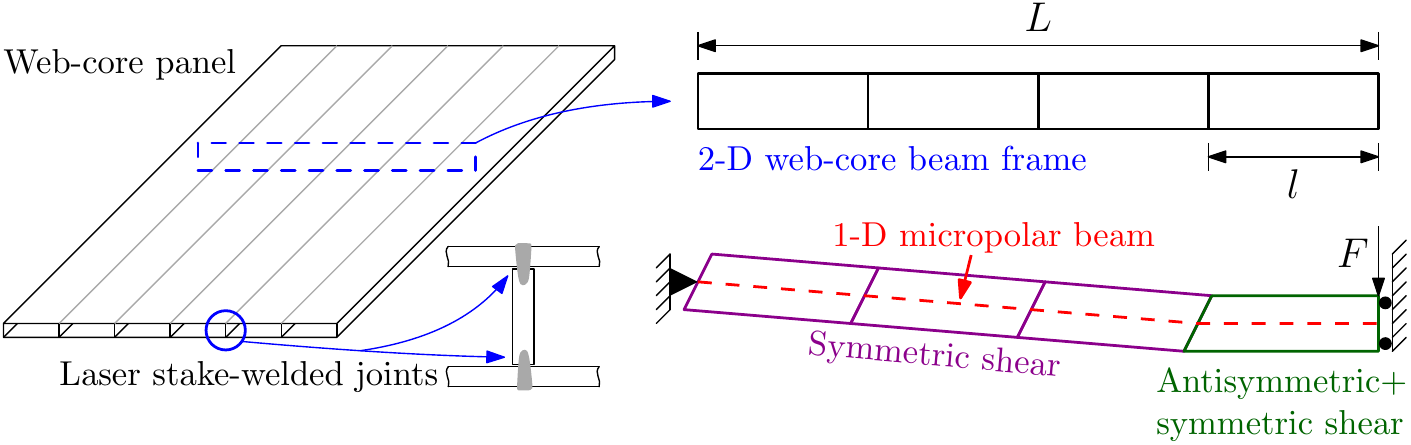}
\caption{The structure studied here represents a plane section of a laser-welded web-core sandwich panel. The 2-D web-core beam frame is modeled as a 1-D micropolar beam that allows antisymmetric shear deformation to emerge at locations where the 2-D deformations cannot be reduced to 1-D by considering only symmetric shear behavior.}
\end{figure}

The two-scale approach has its roots in the method first presented by Noor and Nemeth \cite{noor1980,noor1988,ostoja2002}. They derived a micropolar-type Timoshenko beam model by a variational method using the strain and kinetic energy expressions of different rigid-jointed lattice configurations. We operate with a similar strain energy expression here to derive the constitutive relations for the ESL-FSDT Timoshenko beam model developed earlier in the general context of micropolar elasticity \cite{karttunen2018a}.

In a more contemporary vein, the present two-scale constitutive modeling method is based on ideas that are comparable to those behind second-order computational homogenization techniques \cite{kouznetsova2002,larsson2007,geers2010,matouvs2017}: (1) No constitutive model is assumed for the macroscale beam a priori. (2) A microscale unit cell of the beam is modeled by classical constituents (i.e., by conventional beam elements). (3) The macroscale beam kinematics are imposed on the microscale unit cell in order to bridge the two scales. However, instead of solving a nested boundary value problem as in computational homogenization, the (hyperelastic) constitutive relations in the present linear case will be determined directly from the unit cell strain energy given in terms of the macroscale beam strains. In a broad sense, the two-scale approach for a 1-D micropolar beam is a step towards a general constitutive modeling technique in micropolar elasticity with particular emphasis on mid-surface structural components such as beams, plates and shells made of lattice materials. The micropolar modeling of lattices does not usually employ mid-surface kinematics (see, e.g. \cite{kumar2004,spadoni2012,trovalusci2017}).

The rest of the paper is organized as follows. The micropolar Timoshenko beam model \cite{karttunen2018a} is briefly reviewed in Section 2. The two-scale constitutive modeling of a web-core beam which gives the stress resultant equations for the beam is carried out in Section 3. A novel micropolar Timoshenko beam finite element is formulated in Section 4 on the basis of a general static displacement solution to the beam equilibrium equations. We also provide new means for the calculation of the periodic classical stress response of 2-D web-core frames from 1-D micropolar beam solutions. Numerical examples are presented in Section 5 and concluding remarks in Section 6.
\section{Micropolar Timoshenko beam theory}
The pertinent equations of a 1-D micropolar Timoshenko beam model \cite{karttunen2018a} are reviewed here with the axial behavior of the beam taken into account as a new contribution. We consider a beam of length $L$ and height $h$, as shown in Fig.~2. The displacements $U_x$ and $U_y$ and the microrotation $\Psi$ of the beam can be expressed in terms of the central axis kinematic variables ($u_x, u_y, \phi, \psi$) as
\begin{equation}
U_x(x,y)=u_x(x)+y\phi(x), \quad U_y(x,y)=u_y(x), \quad \Psi(x,y)=\psi(x),
\end{equation}
where $u_x$ is the axial displacement, $\phi$ is the rotation of the cross-section, $u_y$ is the transverse deflection, and $\psi$ is an independent microrotation that will ultimately describe the rotation of the flexible joints of the web-core beam.
\begin{figure}
\centering
\includegraphics[scale=0.9]{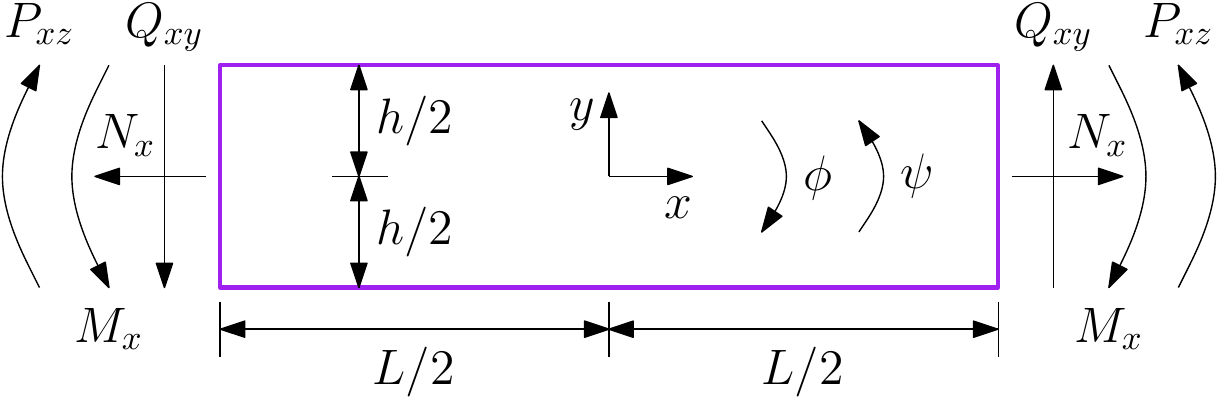}
\caption{Micropolar Timoshenko beam. The positive directions of the stress resultants and rotations are shown.}
\end{figure}
The nonzero strains of the beam are
\begin{equation}
\begin{aligned}
\epsilon_x&=\frac{\partial U_x}{\partial x}=u_x'+y\phi'=\epsilon_x^0+y\kappa_{x}, \ & \kappa_{xz}&=\frac{\partial \Psi}{\partial x}=\psi' \\ \epsilon_{xy}&=\frac{\partial U_y}{\partial x}-\Psi=u_y'-\psi, \ & \epsilon_{yx}&=\frac{\partial U_x}{\partial y}+\Psi=\phi+\psi,
\end{aligned}
\end{equation}
where the prime ``$'$" on the variables denotes differentiation with respect to $x$. The curvature $\kappa_{xz}$ describes the bending of the face sheets of the web-core beam with respect to their own centroid axes. The symmetric and antisymmetric shear strains of the beam are defined as
\begin{align}
\gamma_s&=\epsilon_{xy}+\epsilon_{yx}=u_y'+\phi, \\ \gamma_a&=\epsilon_{xy}-\epsilon_{yx}=u_y'-\phi-2\psi=2(\omega_z-\psi),
\end{align}
respectively, where $\omega_z$ is the macrorotation. The symmetric part takes the same form as the shear strain in the classical Timoshenko beam theory. The antisymmetric part is defined by the difference between the macrorotation and the microrotation. Evidently, for $\omega_z=\psi$ we have $\gamma_a=0$  and the relative strains reduce to their classical definitions \cite{barber2010}, for example, $\epsilon_{xy}=u_y'-\omega_z=(u_y'+\phi)/2$.

In addition to having an independent rotational degree of freedom, the micropolar beam can transmit couple-stress $m_{xz}$, as well as the usual  stresses $\sigma_x, \tau_{xy}$, and $\tau_{yx}$. The four equilibrium equations of the beam can be written in terms of the stress resultants ($N_x, M_x, Q_{xy}, Q_{yx}, P_{xz}$) as
\begin{align}
N_x'=0, \quad M_x'-Q_{yx}=0, \quad Q_{xy}'=-q, \quad
P_{xz}'+Q_{xy}-Q_{yx}=-m,
\end{align}
where $q$ is a distributed transverse load and $m$ is a distributed externally applied couple. The shear forces are not necessarily equal (i.e., $Q_{xy}\neq Q_{yx}$). As for the boundary conditions, one element in each of the following four duality pairs should be specified at the beam ends
\begin{equation}
N_x \quad \textrm{or} \quad u_x, \quad  Q_{xy} \quad \textrm{or} \quad u_y, \quad M_x \quad \textrm{or} \quad \phi, \quad  P_{xz} \quad \textrm{or} \quad \psi.
\end{equation}
The symmetric and antisymmetric shear forces are defined as
\begin{align}
Q_s&=\frac{Q_{xy}+Q_{yx}}{2}, \\
Q_a&=\frac{Q_{xy}-Q_{yx}}{2},
\end{align}
respectively (see Fig.~3). In the following, the use of the symmetric and antisymmetric shear strains and forces facilitates the determination of the constitutive relations that complete the micropolar Timoshenko beam theory.
\begin{figure}
\centering
\includegraphics[scale=1]{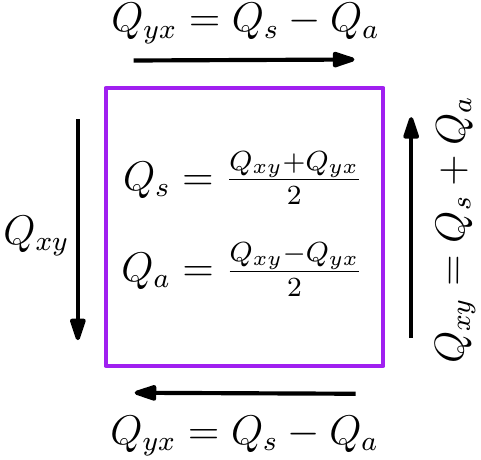}
\caption{Split of the shear forces into symmetric and antisymmetric parts.}
\end{figure}
\section{Two-scale constitutive modeling}
Figure 4 shows, for modeling purposes, a web-core unit cell attached to an arbitrary cross section of the micropolar beam. The micropolar beam of length $L$ is a \textit{macrostructure} and the unit cell of length $l$ represents its periodic \textit{microstructure}. In order to obtain the constitutive equations for the beam model, the strain energy of the microscale unit cell needs to be expressed in terms of the macroscale strains (2)--(4). This is achieved in the next sections through continualization of the unit cell corner displacements in combination with finite element modeling of the unit cell by classical constituents (i.e., Euler--Bernoulli beam elements). The resulting constitutive matrix of the beam is positive definite, which implies that it describes a stable lattice material.
\begin{figure}
\centering
\includegraphics[scale=1.17]{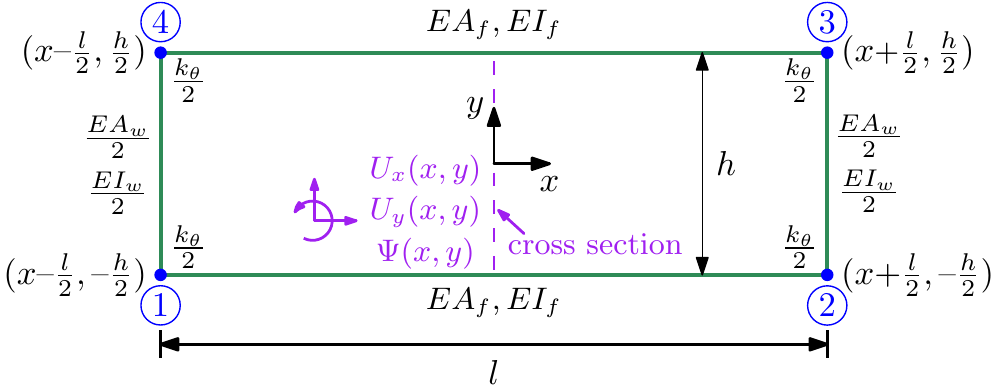}
\caption{Arbitrary cross section of the micropolar beam of length $L$ possessing microstructure of length $l$. The displacements at nodes 1 to 4 of the web-core microstructure are expressed in terms of the cross-sectional beam displacements and micropolar strains via a 1-D Taylor series expansion.}
\end{figure}
\subsection{Continualization of unit cell displacements}
The unit cell corner displacements in Fig.~4 are expressed in terms of the cross-sectional displacements $U_x$ and $U_y$ and rotation $\Psi$. With distance from an arbitrary cross section located within the interval $-L/2\leq x\leq L/2$, Taylor series expansions of Eqs.~(1) lead to
\begin{align}
U_x(x\pm l/2,\pm h/2)&= u_x\pm\frac{h}{2}\left[\frac{1}{2}(\gamma_s-\gamma_a)-\psi\right]\pm\frac{l}{2}\left(\epsilon_x^0\pm\frac{h}{2}\kappa_x\right) , \\
U_y(x\pm l/2,\pm h/2)&=u_y\pm\frac{l}{2}\left[\frac{1}{2}(\gamma_s+\gamma_a)+\psi\right] , \\
\Psi(x\pm l/2,\pm h/2)&=\psi\pm\frac{l}{2}\kappa_{xz} ,
\end{align}
where the micropolar strains (2)--(4) have been imposed on the cross-sectional rotation $\phi$ and the displacement gradients. Using the node numbering of Fig.~4, we can write the discrete-to-continuum transformation given by Eqs.~(9)--(11) in matrix form
\begin{equation}
\mathbf{d}=\mathbf{T}^{\phantom{ }}_u\mathbf{u}+\mathbf{T}^{\phantom{ }}_\epsilon\boldsymbol{\epsilon},
\end{equation}
where the generalized discrete displacement vector is
\begin{equation}
\mathbf{d}=\left\{U_{x,1} \ \ U_{y,1} \ \ \Psi_{1} \ \ U_{x,2} \ \ U_{y,2} \ \ \Psi_{2} \ \ U_{x,3} \ \ U_{y,3} \ \ \Psi_{3} \ \ U_{x,4} \ \ U_{y,4} \ \ \Psi_{4}\right\}^{\textrm{T}}
\end{equation}
and the vectors for the continuous variables read
\begin{align}
\mathbf{u}&=\left\{u_x \ \ u_y \ \ \phi \ \ \psi\right\}^{\textrm{T}} , \\
\boldsymbol{\epsilon}&=\{\epsilon_x^0 \quad \kappa_x \quad \gamma_s \quad \gamma_a \quad \kappa_{xz}\}^{\textnormal{T}} .
\end{align}
The transformation matrices $\mathbf{T}^{\phantom{ }}_u$ and $\mathbf{T}^{\phantom{ }}_\epsilon$ are given in Appendix A. The transformation by Eq.~(12) is not limited to the web-core unit cell but may be applied to any lattice unit cell that can be reduced to the four-node presentation of Fig.~4 through static condensation.
\subsection{Unit cell strain energy and beam constitutive matrix}
The web-core microstructure displayed in Fig.~4 can be modeled by using four nodally-exact Euler--Bernoulli or Timoshenko beam finite elements as both choices result in a system that is consistent with the generalized displacement vector (13). As for their material composition, the elements can be heterogeneous and anisotropic. However, the web-core structure at hand (Fig.~1) consists of relatively thin components made of steel, so we use linearly elastic isotropic, homogeneous Euler--Bernoulli beam elements in this study.

The web-core unit cell in Fig.~4 has similar top and bottom faces with axial stiffness $EA_f$ and bending stiffness $EI_f$. Only half of each web stiffness parameter is accounted for due to symmetry between neighboring unit cells so that we have $EA_w/2$, $EI_w/2$ and $k_\theta/2$ for the axial, bending and joint stiffnesses, respectively. While the faces are modeled using normal Euler--Bernoulli beam elements, the webs are modeled using special-purpose elements with rotational springs at both ends to account for the flexibility of the laser-welded joints \cite{monforton1963,chen2005,romanoff2007c}.

The strain energy of the web-core unit cell modeled by using four Euler--Bernoulli beam finite elements can be written as
\begin{equation}
W=\frac{1}{2}\mathbf{d}^{\textrm{T}}\mathbf{k}\mathbf{d},
\end{equation}
where $\mathbf{k}$ is the global twelve-by-twelve stiffness matrix of the unit cell. By applying the transformation (12) to the strain energy (16) it is straightforward to verify that the displacement terms (14) do not contribute to the strain energy and we obtain
\begin{equation}
W=\frac{1}{2}\boldsymbol{\epsilon}^{\textrm{T}}\mathbf{T}_\epsilon^{\textrm{T}}\mathbf{k}\mathbf{T}^{\phantom{ }}_\epsilon\boldsymbol{\epsilon}.
\end{equation}
We define the \textit{linear density} of the unit cell strain energy as
\begin{equation}
W_0^l\equiv\frac{W}{l}=\frac{1}{2}\boldsymbol{\epsilon}^{\textrm{T}}\mathbf{C}\boldsymbol{\epsilon}
\end{equation}
where the constitutive matrix is given by
\begin{equation}
\mathbf{C}=\frac{1}{l}\mathbf{T}_\epsilon^{\textrm{T}}\mathbf{k}\mathbf{T}^{\phantom{ }}_\epsilon.
\end{equation}
The unit cell represents a \textit{lattice material} of which the micropolar beam is made of. Therefore, in analogy with any hyperelastic material, we write for the micropolar beam continuum
\begin{equation}
\mathbf{S}\equiv\frac{\partial W_0^l}{\partial \boldsymbol{\epsilon}}=\mathbf{C}\boldsymbol{\epsilon},
\end{equation}
where $\mathbf{S}$ is now the stress resultant vector of the micropolar beam. The explicit form of Eq.~(20) is
\begin{equation}
\begin{Bmatrix}
N_x \\
M_x \\
Q_s \\
Q_a \\
P_{xz}
\end{Bmatrix}
=
\begin{bmatrix}
 2 EA_f & 0 & 0 & 0 & 0 \\
  & \frac{EA_f h^2}{2}+\Theta  & 0 & 0 & \Theta  \\
  &  & \frac{6EI_f+\Theta}{l^2} & \frac{6EI_f-\Theta}{l^2} & 0 \\
  & \textrm{SYM}  &  & \frac{6EI_f+\Theta}{l^2} & 0 \\
  & &  &  & 2EI_f +\Theta \\
\end{bmatrix}
\begin{Bmatrix}
\epsilon_x^0 \\
\kappa_x \\
\gamma_s \\
\gamma_a \\
\kappa_{xz}
\end{Bmatrix}
\end{equation}
where
\begin{equation}
\Theta=\frac{3EI_w k_\theta l}{6EI_w+k_\theta h}.
\end{equation}
By using the constitutive relations (21) the equilibrium equations (5) can be solved for the displacements. Finally, in the micropolar theory it holds that
\begin{equation}
\tau_{xy}\epsilon_{xy}+\tau_{yx}\epsilon_{yx}=\tau_{s}\gamma_{s}+\tau_{a}\gamma_{a}
\end{equation}
from which it follows that the strain energy of the micropolar beam can be written as \cite{karttunen2018a}
\begin{equation}
\begin{aligned}
U&=\frac{1}{2}\int_V(\sigma_x\epsilon_x+\tau_{s}\gamma_{s}+\tau_{a}\gamma_{a}+m_{xz}\kappa_{xz})dV \\
&=\frac{1}{2}\int_{-L/2}^{L/2}\left(N_x\epsilon_x^0+M_x\kappa_x+Q_{s}\gamma_s+Q_{a}\gamma_a+P_{xz}\kappa_{xz}\right)dx \\
&=\frac{1}{2}\int_{-L/2}^{L/2}\left(\boldsymbol{\epsilon}^{\textrm{T}}\mathbf{C}\boldsymbol{\epsilon}\right)dx,
\end{aligned}
\end{equation}
which shows that the linear density of the strain energy of the micropolar beam is the same as that of the unit cell [see Eq.~(18)]. This result underscores the fact that the determination of the constitutive equations (20) and the bridging of the two scales is founded on an assumption of strain energy equivalence between the macrostructure (beam) and the microstructure (unit cell).
\subsection{Validity of the constitutive matrix}
The constitutive matrix (19) represents a lattice material in the context of the micropolar beam theory. In order for the material to be stable in the conventional sense ($U>0$ for nonzero $\boldsymbol{\epsilon}$), the constitutive matrix $\mathbf{C}$ should be positive definite \cite{ting1996}. The matrix is positive definite if all (1) eigenvalues or, equivalently, (2) leading principal minors of the matrix are positive and nonzero. The symbolic expressions for the leading principal minors are considerably simpler:
\begin{equation}
\begin{aligned}
C^1&=|C_{11}|=2EA_f, \\
C^2&=
\begin{vmatrix}
C_{11} & C_{12} \\
C_{12} & C_{22}
\end{vmatrix}=EA_f(EA_fh^2+2\Theta)  \\
C^3&=\ldots=EA_f(6EI_f+\Theta)(EA_fh^2+2\Theta)/l^2  \\
C^4&=\ldots=24EA_fEI_f\Theta(EA_fh^2+2\Theta)/l^4 \\
C^5&=|\mathbf{C}|=24EA_fEI_f\Theta\left[4EI_f\Theta+EA_fh^2(2EI_f+\Theta)\right]/l^4.
\end{aligned}
\end{equation}
Because all beam parameters are positive, we have
\begin{equation}
C^k>0 \quad \textrm{for} \ k=1,2,\ldots,5
\end{equation}
and the constitutive matrix (19) represents a stable lattice material. It may also be stated that Eq.~(26) guarantees the stability of the thermodynamic state of the micropolar beam \cite{eringen2012}.
\subsection{Interpretation of the web-core stiffness parameters}
\begin{itemize}
\item In the constitutive matrix (19), $C_{11}=2EA_f$ is the axial stiffness due to the two faces, whereas $EA_fh^2/2$ under $C_{22}$ is the usual global bending stiffness generated by the sandwich effect \cite{allen1969}. If the top and bottom face sheets had different thicknesses, a coupling term $C_{12}$ would appear in the matrix. The axial web stiffness $EA_w$ does not appear in the constitutive matrix at all because of the transverse inextensibility of the beam $(U_y=u_y)$.
\item The shear stiffnesses $C_{33}$, $C_{34}$ and $C_{44}$ account for the shear behavior of the micropolar Timoshenko beam model as well as possible through the beam-cell energy equivalence scheme. The model does not employ any extrinsic micropolar shear correction factors. 
\item The bending moments $M_x$ and $P_{xz}$ are coupled by $C_{25}=\Theta$ and the shear forces $Q_s$ and $Q_a$ by $C_{34}$. Some lattice materials may have fully populated constitutive matrices $\mathbf{C}$ and exhibit stronger (anisotropic) coupling; however, such materials are not studied in this paper.
\item The laser-welded joints of the web-core beam are rigid for infinite rotational stiffness $k_\theta$ and pinned for zero rotational stiffness. For these two limiting cases we have
\begin{equation}
k_\theta\rightarrow\infty:\ \Theta=\frac{3EI_wl}{h} \quad \textrm{and} \quad k_\theta\rightarrow0:\ \Theta=0
\end{equation}
In the case of pinned joints $(\Theta=0)$, the coupling between the bending moments $M_x$ and $P_{xz}$ vanishes and the sandwich effect is also lost so that the beam is essentially an Euler--Bernoulli beam with bending stiffness equal to the local bending stiffness $2EI_f$.
\end{itemize}
\section{Finite element based on general displacement solution}
\subsection{General displacement solution}
In order to derive a general solution to the beam equilibrium equations (5) which can be used as the basis for finite element formulations, we substitute the constitutive matrix
\begin{equation}
\mathbf{C}=
\begin{bmatrix}
C_{11} & C_{12} & 0 & 0 & C_{15} \\
 & C_{22} & 0 & 0 & C_{25} \\
 &  & C_{33} & C_{34} & 0 \\
 & \textrm{SYM} &  & C_{44} & 0 \\
 &  &  &  & C_{55}
\end{bmatrix}
\end{equation}
into the constitutive equations (21). The coupling terms $C_{12}$ and $C_{15}$, although not needed for the web-core beam (see Fig.~1), do not complicate the analytical solution substantially; further additional terms would. A brief outline of the solution process is given in Appendix B. The homogeneous solution ($q=m=0$) to the equilibrium equations (5) is
\begin{align}
u_x&=c_1+c_2x+\frac{C_{15}-C_{12}}{C_{11}}c_6x^2+\alpha_1\left(c_7\textrm{e}^{\beta_1 x}+c_8\textrm{e}^{-\beta_1 x}\right), \\
u_y&=c_3-c_4x-c_5\frac{x^2}{2}-c_6\left(\frac{x^3}{3}-\beta_3x\right)-\frac{\beta_2}{\beta_1}\left(c_7\textrm{e}^{\beta_1 x}-c_8\textrm{e}^{-\beta_1 x}\right), \\
\phi&=c_4+c_5x+c_6x^2+c_7\textrm{e}^{\beta_1 x}+c_8\textrm{e}^{-\beta_1 x},\\
\psi&=-c_4-c_5x-c_6(x^2-\alpha_3)+\alpha_2\left(c_7\textrm{e}^{\beta_1 x}+c_8\textrm{e}^{-\beta_1 x}\right)
\end{align}
and constants $\alpha_j$ and $\beta_j$ ($j=1,2,3$) are given in Appendix B, as well as particular solutions for uniformly distributed pressure and moment loads $q=q_0$ and $m=m_0$, respectively. In the homogeneous solution, the integration constants $c_1$, $c_3$ and $c_4$ correspond to rigid body motions and the five remaining constants are related to the stress resultants (cf.~\cite{karttunen2016c}).
\subsection{Shape functions for micropolar Timoshenko beam element}
The general analytical solution (29)--(32) containing polynomial and exponential terms is used for the derivation of a nodally-exact beam finite element without introducing any shape function approximations. Fig.~5 presents the setting according to which the element is developed. Both nodes have four degrees of freedom, namely, the axial and transverse displacements $u_{i}$ and $w_{i}$, respectively, and rotations $\phi_{i}$ and $\psi_{i}$ ($i=1,2$). Using Eqs.~(29)--(32), we define the FE degrees of freedom as
\begin{equation}
\begin{aligned}
        u_{1}&=u_x(-L_e/2), & \qquad u_{2}&=u_x(L_e/2), \\
        w_{1}&=u_y(-L_e/2), & \qquad w_{2}&=u_y(L_e/2), \\
        \phi_1&=-\phi(-L_e/2), & \qquad \phi_2&=-\phi(L_e/2) \\
        \psi_1&=\psi(-L_e/2), & \qquad \psi_2&=\psi(L_e/2).
\end{aligned}
\end{equation}
In matrix form we have
\begin{equation}
\mathbf{\Delta}=\mathbf{H}\mathbf{c},
\end{equation}
where the generalized micropolar displacement vector is
\begin{equation}
\mathbf{\Delta}=\left\{u_1 \ \ w_1 \ \ \phi_1 \ \ \psi_1 \ \ u_2 \ \ w_2 \ \ \phi_2 \ \ \psi_2 \right\}^{\textrm{T}}
\end{equation}
and $\mathbf{H}$ is a coefficient matrix and $\mathbf{c}$ contains the constant coefficients $c_i$ ($i=1,\ldots,8$), which are obtained in terms of the FE degrees of freedom by
\begin{equation}
\mathbf{c}=\mathbf{H}^{-1}\mathbf{\Delta}.
\end{equation}
The kinematic variables (29)--(32) in terms of the FE degrees of freedom may then be written as
\begin{align}
\mathbf{u}=
\begin{Bmatrix}
u_x(x) \\
u_y(x) \\
\phi(x)\\
\psi(x)
\end{Bmatrix}
=\mathbf{A}\mathbf{c}=\mathbf{A}\mathbf{H}^{-1}\mathbf{\Delta}=
\begin{Bmatrix}
\mathbf{N}_{ux} \\
\mathbf{N}_{uy} \\
\mathbf{N}_{\phi} \\
\mathbf{N}_{\psi}
\end{Bmatrix}\mathbf{\Delta}=\mathbf{N}_u\mathbf{\Delta},
\end{align}
where $\mathbf{A}$ is a matrix with polynomial and exponential terms and the four-by-eight matrix $\mathbf{N}_u$ contains the shape functions. The general solution and the formulation of the shape functions are given in an online supplementary Mathematica file \textit{MicropolarShapeFunctions}. 
\begin{figure}
\centering
\includegraphics[scale=0.75]{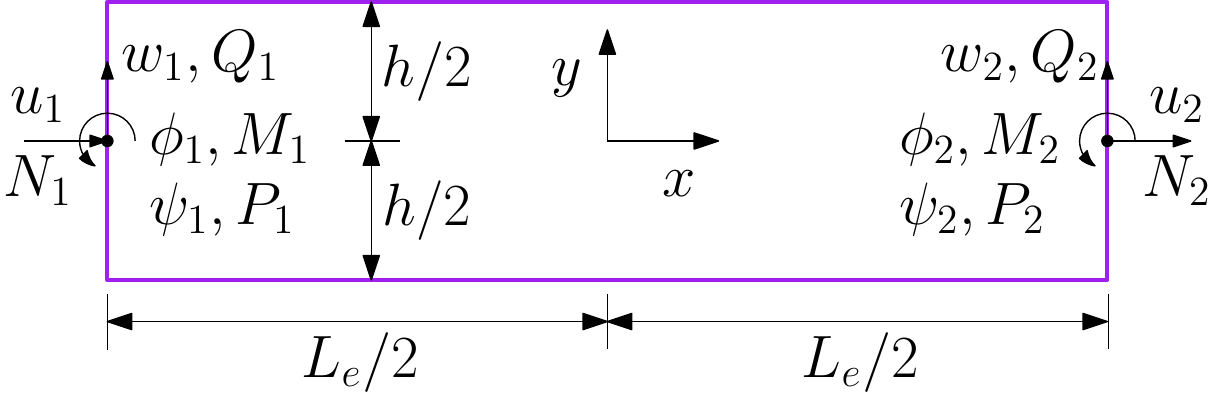}
\caption{Set-up according to which the micropolar Timoshenko beam finite element is developed.}
\end{figure}
\subsection{Kinetic energy for consistent mass matrix}
In addition to static applications, the shape functions (37) can be used for the derivation of a consistent mass matrix for the beam. To obtain the kinetic energy of the micropolar beam for that purpose, only the constant terms are included in the expansions (9)--(11). Then the discrete-to-continuum transformation of the kinetic energy of the unit cell can be written as
\begin{equation}
\Tilde{K}=\frac{1}{2}\mathbf{\dot{d}}^{\textrm{T}}\mathbf{\Tilde{m}}\mathbf{\dot{d}}=\frac{1}{2}\mathbf{\dot{u}}^{\textrm{T}}\mathbf{T}_{\dot{u}}^{\textrm{T}}\mathbf{\Tilde{m}}\mathbf{T}^{\phantom{ }}_{\dot{u}}\mathbf{\dot{u}},
\end{equation}
where the dot on the variables denotes differentiation with respect to time. The global twelve-by-twelve mass matrix $\mathbf{\Tilde{m}}$ of the unit cell is modeled by using Euler--Bernoulli beam elements with consistent mass matrices. The transformation matrix $\mathbf{T}^{\phantom{ }}_{\dot{u}}$ is given in Appendix A. We define the linear density of the unit cell kinetic energy as
\begin{equation}
\Tilde{K}_0^l\equiv\frac{\Tilde{K}}{l}=\frac{1}{2}\mathbf{\dot{u}}^{\textrm{T}}\mathbf{m}\mathbf{\dot{u}}
\end{equation}
where
\begin{equation}
\mathbf{m}=\rho
\begin{bmatrix}
 2 A_f+\frac{h  A_w}{l} & 0 & 0 & 0 \\
 0 & 2  A_f+\frac{h  A_w}{l} & 0 & 0 \\
 0 & 0 & \frac{h^2 (70 l  A_f+17 h  A_w)}{140 l} & \frac{3 h^3  A_w}{140 l} \\
 0 & 0 & \frac{3 h^3  A_w}{140 l} & \frac{ A_w h^3+2 l^3  A_f}{210 l} \\
\end{bmatrix}.
\end{equation}
Then we assume that the linear densities of the beam and unit cell kinetic energies are equal
\begin{equation}
K_0^l\equiv\Tilde{K}_0^l.
\end{equation}
Finally, total kinetic energy of the beam is
\begin{equation}
K=\int_{-L/2}^{L/2}K_0^l\ dx=\frac{1}{2}\int_{-L/2}^{L/2}\mathbf{\dot{u}}^{\textrm{T}}\mathbf{m}\mathbf{\dot{u}}\ dx.
\end{equation}
\subsection{Finite element equations}
The Lagrangian of the micropolar beam is
\begin{equation}
\mathcal{L}=K-(U-W_d-W_s)
\end{equation}
where $K$ and $U$ are given by Eqs.~(42) and (24), respectively, and the potential energy contribution due to the distributed external loads is
\begin{equation}
W_d=\int_{-L_e/2}^{L_e/2}(qu_y+m\psi)dx
\end{equation}
and the beam end surface tractions bring about the work \cite{karttunen2018a}
\begin{equation}
W_s=\left[N_xu_x+Q_{xy}u_y+M_x\phi+P_{xz}\psi\right]_{-L_e/2}^{L_e/2}\equiv\mathbf{\Delta}^{\textrm{T}}\mathbf{f}
\end{equation}
where
\begin{equation}
\mathbf{f}=\left\{N_1 \ \ Q_1 \ \ M_1 \ \ P_1 \ \ N_2 \ \ Q_2 \ \ M_2 \ \ P_2\right\}^{\textrm{T}}.
\end{equation}
By using the shape function formalism (37) in the Lagrangian (43) [here $\mathbf{\Delta}=\mathbf{\Delta}(t)$], the Lagrange equations
\begin{equation}
\frac{d}{dt\vphantom{\mathbf{\dot{\Delta}}}}\frac{\partial \mathcal{L}}{\partial \mathbf{\dot{\Delta}}}-\frac{\partial \mathcal{L}}{\partial \mathbf{\Delta}\vphantom{\mathbf{\dot{\Delta}}}}=0
\end{equation}
lead to the finite element equations
\begin{equation}
\mathbf{M}\mathbf{\ddot{\Delta}}+\mathbf{K}\mathbf{\Delta} =\mathbf{f}+\mathbf{q}+\mathbf{m}
\end{equation}
where the consistent mass matrix and stiffness matrix are
\begin{equation}
\mathbf{M}=\int_{-L_e/2}^{L_e/2}\mathbf{N}_{u}^{\textrm{T}}\mathbf{m}\mathbf{N}_{u}^{\phantom{}}dx, \quad \mathbf{K}=\int_{-L_e/2}^{L_e/2}\mathbf{N}_{\epsilon}^{\textrm{T}}\mathbf{C}\mathbf{N}_{\epsilon}^{\phantom{}}dx ,
\end{equation}
respectively. The shape function matrix $\mathbf{N}_\epsilon$ is based on Eq.~(15) and is easily formed from $\mathbf{N}_{ux}$, $\mathbf{N}_{uy}$, $\mathbf{N}_{\phi}$ and $\mathbf{N}_{\psi}$ and their derivatives, see Eq.~(37). The distributed loads are given by
\begin{equation}
\mathbf{q}+\mathbf{m}=\int_{-L_e/2}^{L_e/2}(q\mathbf{N}_{uy}^{\textrm{T}}+m\mathbf{N}^{\textrm{T}}_{\psi})dx.
\end{equation}
Once the nodal displacements have been calculated by using Eq.~(48), the central axis displacements are obtained from Eq.~(37) and the micropolar in-plane displacement field from Eq.~(1). Then the micropolar response may be localized to obtain the periodic web-core response.

\subsection{Localization by beam elements}
After a 1-D micropolar beam problem has been solved, the corresponding 2-D periodic response is obtained by mapping the micropolar results into the classical Euler--Bernoulli beam elements of the web-core unit cell of Fig.~4. The used Euler--Bernoulli elements are based on a setup similar to that in Fig.~5 and each element has a local coordinate system $\hat{x}-\hat{y}$. For example, for the axial and transverse deflections of a unit cell's top face under a uniformly distributed load $q_0$ we have
\begin{align}
\hat{u}_{x,f}&=\mathbf{\hat{N}}_{ux}\mathbf{\hat{\Delta}}=
\begin{Bmatrix}
 \frac{1}{2}-\frac{\hat{x}}{l} \\
 \frac{1}{2}+\frac{\hat{x}}{l}
\end{Bmatrix}^{\textrm{T}}
\begin{Bmatrix}
U_{x,4} \\
U_{x,3}
\end{Bmatrix}
, \\
\hat{u}_{y,f}&=\mathbf{\hat{N}}_{uy}\mathbf{\hat{\Delta}}=
\begin{Bmatrix}
 \frac{(l+\hat{x})(l-2 \hat{x})^2 }{2 l^3} \\
 \frac{(l+2 \hat{x})(l-2 \hat{x})^2}{8 l^2} \\
 \frac{(l-\hat{x})(l+2\hat{x})^2}{2 l^3} \\
 \frac{(2 \hat{x}-l)(l+2 \hat{x})^2}{8 l^2}
\end{Bmatrix}^{\textrm{T}}
\begin{Bmatrix}
U_{y,4} \\
\Psi_4 \\
U_{y,3} \\
\Psi_3
\end{Bmatrix}+\frac{q_0\hat{x}^4}{24EI_f} ,
\end{align}
respectively, where the latter satisfies the Euler--Bernoulli beam equation $\hat{u}_{y,f}''''=q_0/EI_f$. The current localization scheme assumes that micropolar solutions contribute only to the homogeneous cubic part of $\hat{u}_y$. The particular, fourth-order load term in Eq.~(52) is independent of micropolar considerations. By substituting the micropolar variables (1) that correspond to the nodal joint displacements into Eqs.~(51) and (52), the periodic response for one web span is obtained. The axial force and the bending moment of the top face for each web span are
\begin{equation}
\hat{N}_{x,f}=EA_f\frac{\partial \hat{u}_{x,f}}{\partial \hat{x}} , \quad
\hat{M}_{x,f}=-EI_f\frac{\partial^2 \hat{u}_{y,f}}{\partial \hat{x}^2} ,
\end{equation}
respectively, the normal stress of interest in practical applications is calculated from
\begin{equation}
\hat{\sigma}_{x,f}=\frac{\hat{N}_{x,f}}{A_f}+\frac{\hat{y}\hat{M}_{x,f}}{I_f},
\end{equation}
from which the peak stresses are calculated by $\hat{y}=\pm t_f/2$.

Finally, we take use of the fact that the position of the microstructure is not fixed along the $x$-axis in the (homogenized) micropolar beam; only the distance $l$ between two webs is predetermined. Thus, we consider a chain of unit cells that moves axially over a beam domain and calculate continuously the maximum face sheet stress in the upper left corner (node 4, $\hat{y}=+t_f/2$) of one of the unit cells. This calculation also requires the displacements of the upper right corner (node 3). Ultimately, this gedanken experiment leads to the conclusion that the stress is given by
\begin{align}
\hat{\sigma}_{x,f}^{\textrm{env}}&=\frac{E_f}{l}\left\{u_x(x+l)-u_x(x)+(h/2)\left[\phi(x+l)-\phi(x)\right]\right\}\nonumber \\&+\frac{E_f t_f}{l^2}\left\{3\left[u_y(x)-u_y(x+l)\right]+l\left[\psi(x+l)+2\psi(x)\right]\right\}-\frac{3q_0l^2}{4bt_{f}^{2}},
\end{align}
which can be calculated directly from a 1-D micropolar solution and gives us a continuous envelope curve within the beam domain, as will be demonstrated in the next section (Fig.~8). Such envelope curves are very useful for design purposes \cite{romanoff2014b}. The stress is calculated at the joint (node 4) because local peak values of the normal stress $\sigma_{x,f}$ tend to appear at the joints in the periodic response of a web-core beam -- the envelope curve connects these peaks. The relative ease of the above periodic response calculations in the micropolar framework is largely due to the fact that, unlike in classical 1-D context \cite{romanoff2007a}, the joint rotation is included in the analysis through the microrotation.
\section{Numerical examples}
\subsection{General setup}
We study the bending and vibration of web-core beams and panels of different lengths. A typical laser-welded web-core beam is shown in Fig.~6. The faces and webs of the beam are made of steel ($E_f=212$ GPa, $E_w=200$ GPa, $\nu=0.3$, $\rho=7850$ kg/m$^3$). The average rotational stiffness of the laser-welded joints was determined by measurements in Ref.~\cite{romanoff2007c}; for practical purposes in the present context, each joint has a rotational stiffness of 53500 Nm per unit width. Beams of width $b=0.05$ m ($k_\theta=2675$ Nm) and a panel of width $b=1$ m ($k_\theta=53500$ Nm) are studied here. The panel is assumed to be under plane strain conditions that lead to increased Young's moduli, e.g., $E_f\rightarrow E_f/(1-\nu^2)$ for the faces \cite{allen1969}. The face and web thicknesses are $t_f=2.86$ mm and $t_w=3.97$ mm, respectively. The height, i.e, the distance between the face central axes is $h=40\textrm{ mm}+2(t_f/2)$ mm. The web spacing is $l=0.12$ m. 2-D solutions are computed using Euler--Bernoulli FE beam frames modeled by Abaqus (B23 elements); the pins in simply-supported cases are at the central axis of the 2-D frame so that the model corresponds to 1-D cases.
\begin{figure}
\centering
\includegraphics[scale=0.4]{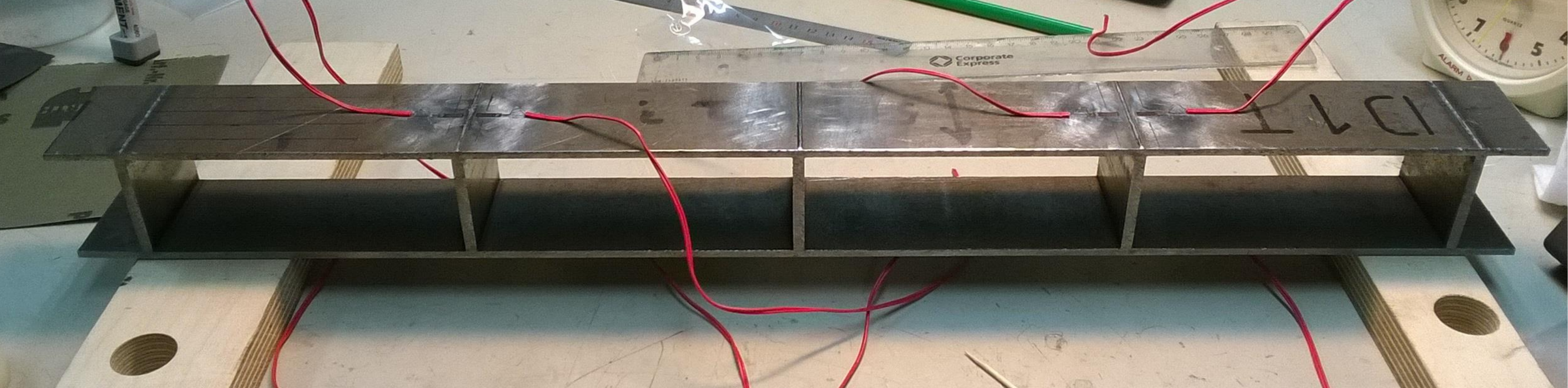}
\caption{Beam-like part cut from a laser-welded sandwich panel and tested for fatigue in Ref.~\cite{karttunen2017b}. The beam is 50 mm wide and four unit cells long ($4l=480$ mm). The web-core represents a \textit{bending-dominated} microstructure \cite{fleck2010}.}
\end{figure}
\subsection{Fixed-fixed beam under a uniformly distributed load}
We model a 50 mm wide fixed-fixed web-core beam by its symmetric half of length $L=6l=0.72$ m. The micropolar beam is under a uniform load $q=-250$ N/m and the boundary conditions are
\begin{equation}
\begin{aligned}
x&=0:\ u_x=u_y=\phi=\psi=0, \\
x&=L:\ u_x=Q_{xy}=\phi=\psi=0,
\end{aligned}
\end{equation}
which are used to solve constants $c_i$ ($i=1,2,\ldots,8$) in Eqs.~(29)-(32). Experimentally the boundary conditions (56) can be realized by a continuous simply-supported beam where each span is loaded by a similar load. Figure 7(a) shows the transverse deflection along the beam, as given by different methods. The 1-D (homogenized) micropolar response is in good agreement with the 2-D finite element beam frame calculations. The localized 1-D face sheet deflection calculated from the micropolar results (cf.\ Section 4.5) is similar to the corresponding 2-D response in its wave-like shape. The classical ESL Timoshenko beam is too flexible and overpredicts the deflection.
\begin{figure}
\centering
\includegraphics[scale=0.63]{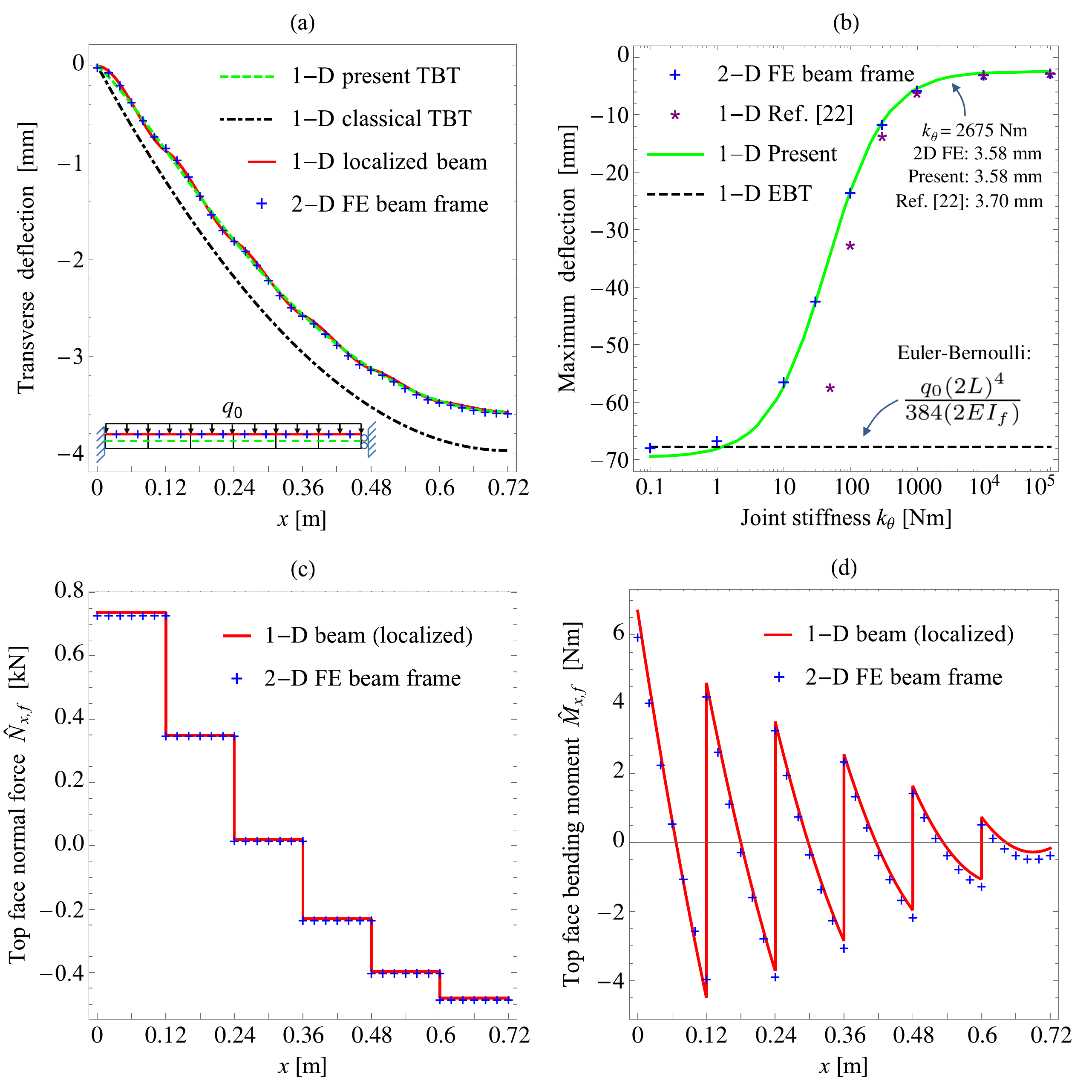}
\caption{Bending response of a fixed-fixed web-core beam under a uniformly distributed load $q=-250$ N/m modeled by a symmetric half. (a) Transverse deflection along the beam. (b) Maximum deflection ($x=0.72$ m) for varying joint stiffness. (c) Localized normal force and (d) bending moment diagrams for the top face of the beam.}
\end{figure}

Figure 7(b) shows the maximum transverse deflection of the web-core beam calculated using different approaches. In Ref.~\cite{karttunen2018a}, only rigid-jointed micropolar web-core beams were considered. When that model is extended to flexible joints, the symmetric and antisymmetric shear stiffnesses of the beam become
\begin{equation}
D_s=D_a=\frac{12}{l}\left(\frac{6l}{h^2EA_f}+\frac{l}{2EI_f}+\frac{h}{EI_w}+\frac{6}{k_\theta}\right)^{-1}.
\end{equation}
Due to similar shear strain definitions, the shear stiffnesses for classical and couple-stress ESL Timoshenko beams are also given by Eq.~(57) \cite{goncalves2017}. Figure 7(b) shows that the early version micropolar model fails to predict the response accurately for flexible joints. This may be due to the fact that the stiffness parameters, including $D_s$ and $D_a$, are determined on the basis of an isotropic micropolar constitutive model \cite{karttunen2018a}. The present 1-D micropolar beam model, which makes no such constitutive assumptions, is in good agreement with 2-D FE results from pinned to rigid joints ($k_\theta:0\rightarrow\infty$). For pinned joints, the 1-D micropolar response is close that of a classical Euler--Bernoulli beam with a total bending stiffness of $2EI_f$. In Fig.~7(b), all models are geometrically linear and the purpose here is to compare some features of models but not to provide physically accurate results for very large displacements.

Figures 7(c) and 7(d) display the periodic normal force and bending moment diagrams for the top face of the beam. The overall correspondence between the localized 1-D results and the 2-D beam frame response is good. The differences between the 1-D and 2-D responses are most notable near the beam supports. In general, it is difficult to impose the boundary conditions and to model the boundary behavior in 1-D exactly in the same way as in 2-D and this may cause some discrepancies between the solutions. Nevertheless, the present micropolar approach that considers antisymmetric shear behavior (Fig.~1) provides a considerable improvement in this respect compared to classical and couple-stress ESL models \cite{karttunen2018a}. Figures 8(a) and 8(b) show the normal stress on the upper surface of the top face calculated from the normal force and bending moment by Eq.~(54). We can see that as the beam becomes longer, the boundary behavior is confined to a smaller area relative to the total beam length and the overall agreement between 1-D and 2-D results improves. The envelope curve provides a simple way to estimate the maximum stresses along the beam. On a practical note, the yield strength of the faces is between 302--322 MPa \cite{romanoff2007c} and the calculated stresses are well below this range.
\begin{figure}
\centering
\includegraphics[scale=0.64]{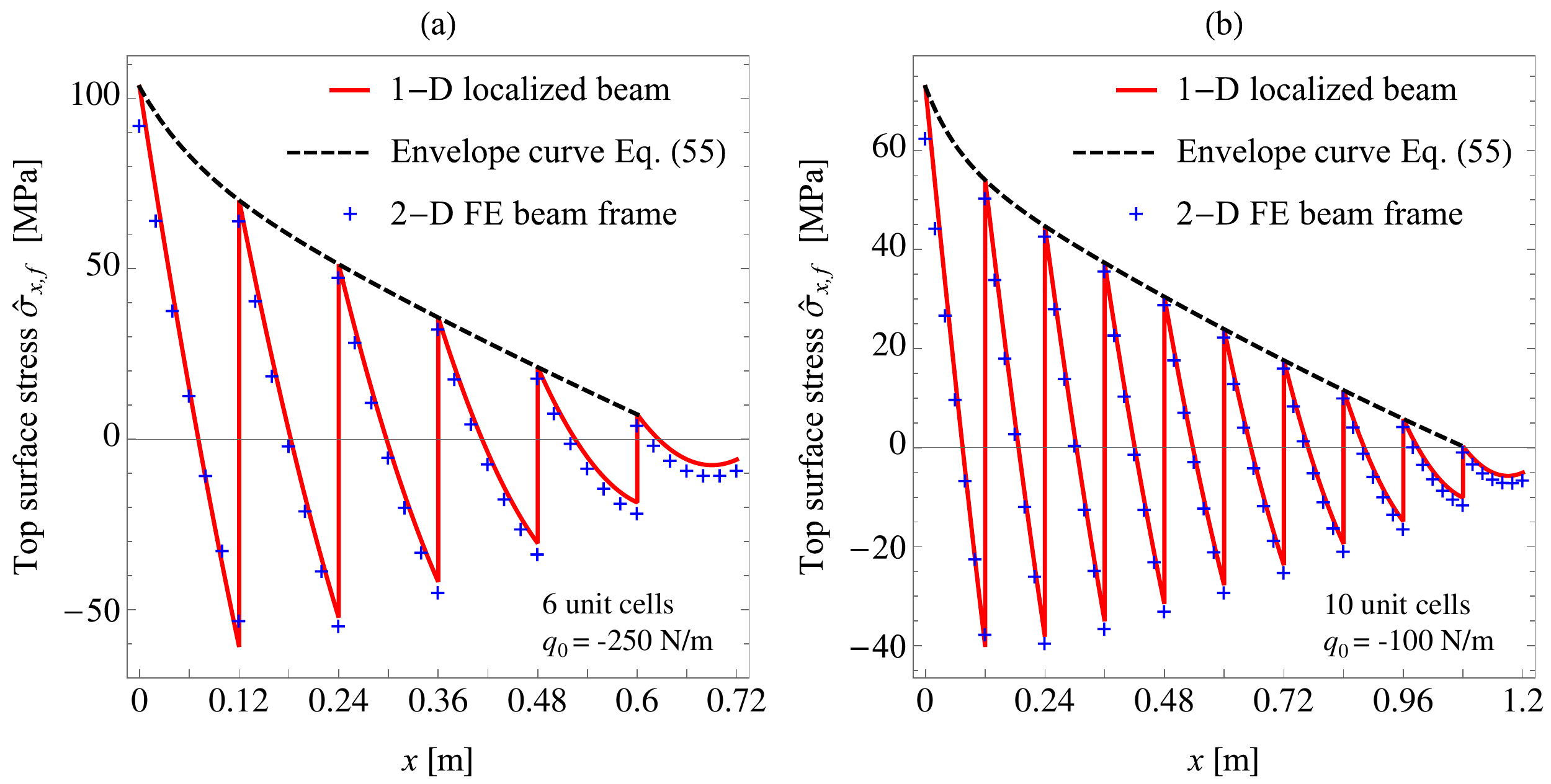}
\caption{Top surface stress of web-core beams of lengths (a) $6l=0.72$ m and (b) $10l=1.2$ m.}
\end{figure}
\subsection{Four-point bending of a web-core panel}
Next we consider a meter-wide panel which is 1.8 meters long (15 unit cells). The panel is in four-point bending and only a symmetric half of it is modeled by using two micropolar beam elements. The in-between node subjected to a vertical point load $F=1000$ N is located at $x=0.6$ m. The constrained degrees of freedom are $u_1=w_1=u_3=\phi_3=\psi_3=0.$
Although the micropolar axial displacement is zero along the beam, the micropolar cross-sectional rotation produces axial displacements in the periodic face sheet response.
\begin{figure}[h!]
\centering
\includegraphics[scale=0.615]{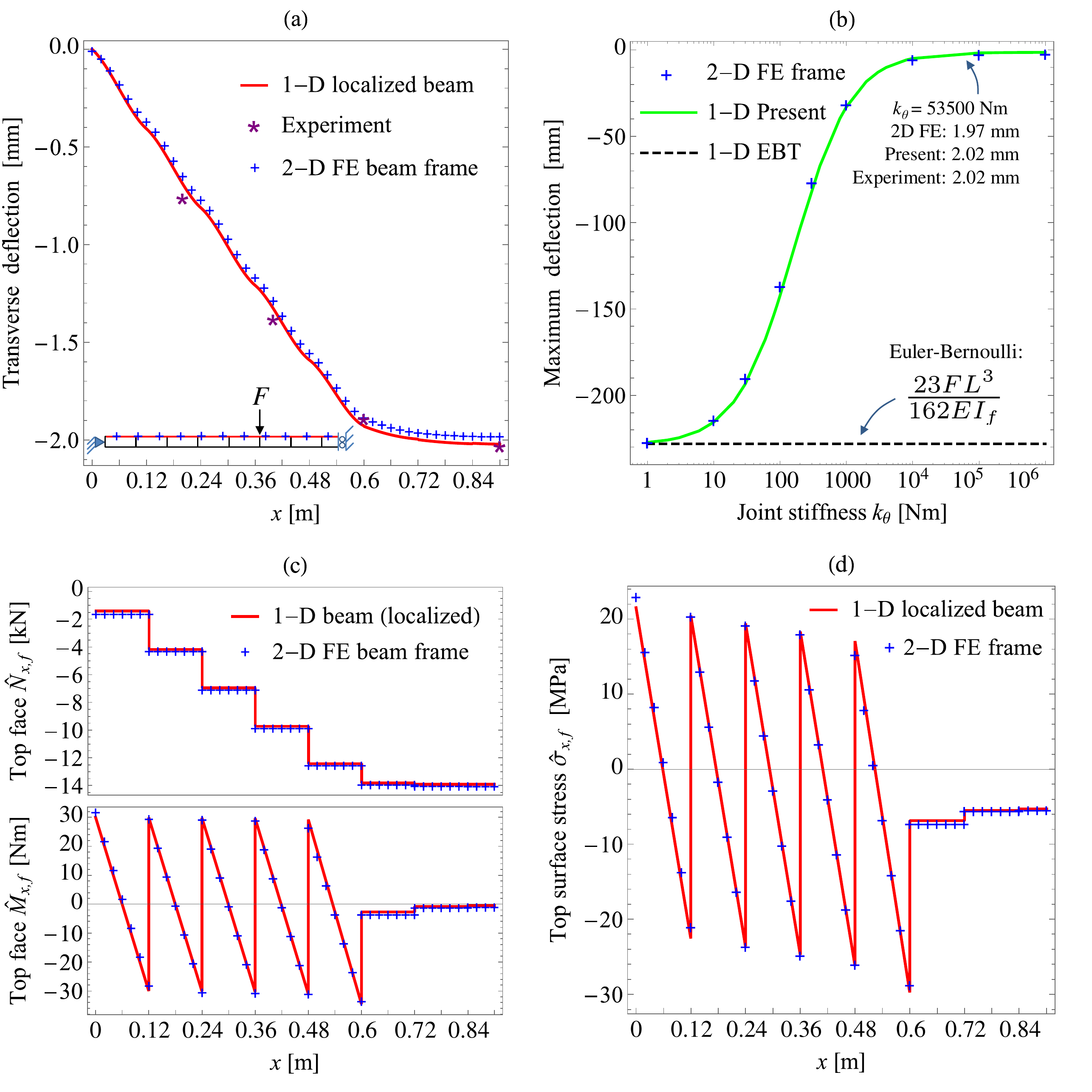}
\caption{Four-point bending of a meter-wide web-core panel modeled by a symmetric half subjected to point load $F=1000$ N. (a) Transverse deflection along the beam. (b) Maximum deflection for varying joint stiffness. (c) Localized normal force and bending moment diagrams for the top face of the beam. (d) Top surface normal stress.}
\end{figure}

Figure 9 is similar to Figs.~7 and 8. Figures 9(a) and 9(b) show that the 1-D micropolar-based periodic displacement response is in good agreement with both experimental \cite{romanoff2007c} and 2-D FE results. Figure 9(c) displays the normal force and bending moment diagrams. In the middle section of the panel where the periodic bending moment is constant, the micropolar transverse deflection does not include a cubic term ($x^3$). Thus, the cubic interpolations have been removed from the localization (52) to obtain constant bending moments. The stress predictions by the 1-D approach in Fig.~9(d) are accurate compared to the 2-D reference solution with only some minor differences in the vicinity of the pinned support and the point load.
\subsection{Natural frequencies of web-core beams}
Figure 10 shows the natural vibration frequencies of cantilever and simply-supported web-core beams calculated using the derived 1-D micropolar beam elements and 2-D FE beam frames modeled by Euler--Bernoulli beam elements. Like in Fig.~6, the beams are 50 mm wide. Both flexible ($k_\theta=2675$ Nm) and rigid joints are considered. The agreement between the 1-D and 2-D results is good for the first eight bending modes. In practical applications, the fundamental frequencies are often of main interest and these are embedded into the figures. For example, for the simply-supported beam the relative errors between the 1-D and 2-D results are around 1\% for the fundamental frequencies.

The present 1-D micropolar beam model can capture the global but not the local bending modes of web-core beams. Examples of both mode types are illustrated in Fig.~11 by 2-D FE beam frame results. In the present case, the local bending modes appear at considerably higher frequencies than the global modes and, thus, are of little practical significance. Further details on the local and global vibrations of laser-welded sandwich plates can be found in a recent paper by Jelovica et al. \cite{jelovica2016}.
\begin{figure}[hb]
\centering
\includegraphics[scale=0.75]{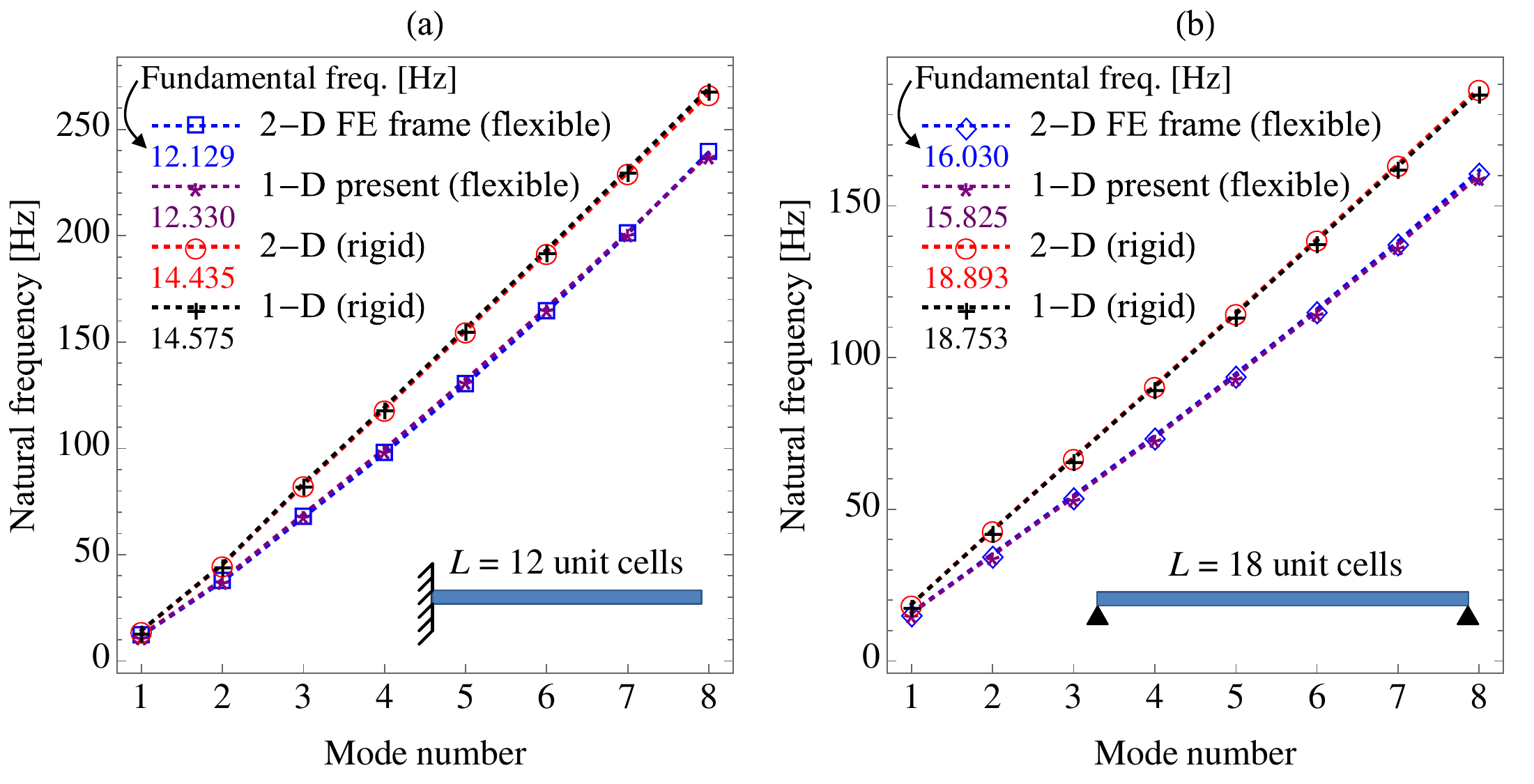}
\caption{Natural vibration frequencies of (a) cantilever and (b) simply-supported web-core beams.}
\end{figure}
\begin{figure}
\centering
\includegraphics[scale=0.48]{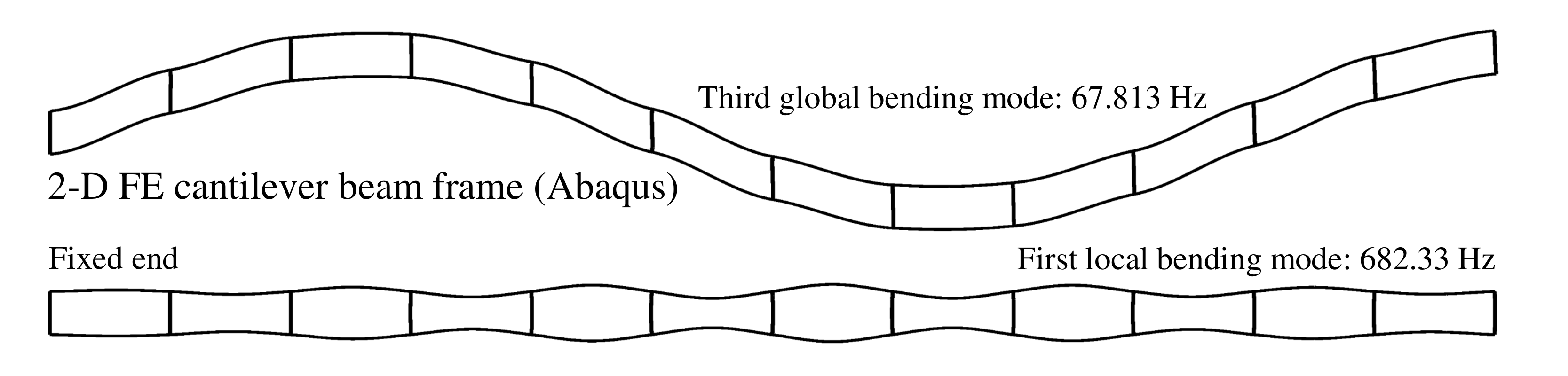}
\caption{Comparison of 2-D global and local bending modes of a cantilever web-core beam with flexible joints.}
\end{figure}
\section{Concluding remarks}
Constitutive equations for a 1-D micropolar Timoshenko beam made of a web-core lattice material were developed by a two-scale energy method. The 1-D micropolar beam gives dimensionally reduced homogenized solutions to 2-D web-core beam frame problems. It was shown that the 1-D micropolar solutions can be localized in a straightforward way to obtain periodic classical stresses for 2-D web-core structures that have moderately flexible joints.

Unlike classical or couple-stress equivalent single layer (ESL) approaches, the micropolar 1-D ESL beam model can capture antisymmetric shear behavior which occurs especially near beam supports and point loads. We stress that the constituents of a 2-D web-core beam frame do not exhibit any antisymmetric shear strains, but when the 2-D problem is reduced to a 1-D ESL beam problem, the antisymmetric behavior needs to be considered.

In recent years, many papers have been published on microstructure-dependent beams that contain an internal length scale as a material parameter (see, e.g., \cite{reddy2007,ma2008,reddy2011,asghari2011}). In these studies it is basically assumed that the microstructure, whatever it may look like, conforms to the chosen constitutive model. The model parameters are usually determined, with adequate success, by fitting model calculations to experimental data without paying heed to the mechanics of the actual microstructure. In contrast to this approach, the present constitutive modeling method for a lattice material did not assume a material model in advance but instead the derivation of the constitutive equations was founded on the actual microstructure. All constitutive parameters emanated from the microstructure at hand. This leads to the conclusion that complex lattice materials can actually have numerous length scale and other material parameters.

The present study dealt with structural components that relate to ship decks, and also to bridges and buildings, highlighting the fact that non-classical continuum mechanics theories have plenty of applications also above the usual nano and micron scale considerations. The applicability of the formulated two-scale constitutive modeling method is determined largely by the periodicity of the structure at hand, not by its size.

\section*{Acknowledgements}
The first author acknowledges that this work has received funding from the European Union's Horizon 2020 research and innovation programme under the Marie Sk\l{}odowska--Curie grant agreement No 745770. The financial support is greatly appreciated. The authors also wish to acknowledge CSC -- IT Center for Science, Finland, for computational resources (Abaqus usage).
\section*{Supplementary file}
The general solution to the governing beam differential equations and the formulation of the shape functions for the beam finite element are given in an online supplementary Mathematica file MicropolarShapeFunctions. 
\appendix
\section{Transformation matrices}
The displacement and strain transformation matrices in Eq.~(12) are
\begin{equation}
\mathbf{T}^{\phantom{ }}_u=
\left[
\begin{array}{cccccccccccc}
 1 & 0 & 0 & 1 & 0 & 0 & 1 & 0 & 0 & 1 & 0 & 0 \\
 0 & 1 & 0 & 0 & 1 & 0 & 0 & 1 & 0 & 0 & 1 & 0 \\
 0 & 0 & 0 & 0 & 0 & 0 & 0 & 0 & 0 & 0 & 0 & 0 \\
 \frac{h}{2} & -\frac{l}{2} & 1 & \frac{h}{2} & \frac{l}{2} & 1 & -\frac{h}{2} & \frac{l}{2} & 1 & -\frac{h}{2} & -\frac{l}{2} &
   1 \\
\end{array}
\right]^{\textrm{T}}
\end{equation}
and
\begin{equation}
\mathbf{T}^{\phantom{ }}_\epsilon=
\left[
\begin{array}{cccccccccccc}
 -\frac{l}{2} & 0 & 0 & \frac{l}{2} & 0 & 0 & \frac{l}{2} & 0 & 0 & -\frac{l}{2} & 0 & 0 \\
 \frac{h l}{4} & 0 & 0 & -\frac{h l}{4} & 0 & 0 & \frac{h l}{4} & 0 & 0 & -\frac{h l}{4} & 0 & 0 \\
 -\frac{h}{4} & -\frac{l}{4} & 0 & -\frac{h}{4} & \frac{l}{4} & 0 & \frac{h}{4} & \frac{l}{4} & 0 & \frac{h}{4} & -\frac{l}{4} &
   0 \\
 \frac{h}{4} & -\frac{l}{4} & 0 & \frac{h}{4} & \frac{l}{4} & 0 & -\frac{h}{4} & \frac{l}{4} & 0 & -\frac{h}{4} & -\frac{l}{4} &
   0 \\
 0 & 0 & -\frac{l}{2} & 0 & 0 & \frac{l}{2} & 0 & 0 & \frac{l}{2} & 0 & 0 & -\frac{l}{2} \\
\end{array}
\right]^{\textrm{T}}
\end{equation}
respectively. The velocity transformation matrix in Eq.~(38) is
\begin{equation}
\mathbf{T}^{\phantom{ }}_{\dot{u}}=\left[
\begin{array}{cccccccccccc}
 1 & 0 & 0 & 1 & 0 & 0 & 1 & 0 & 0 & 1 & 0 & 0 \\
 0 & 1 & 0 & 0 & 1 & 0 & 0 & 1 & 0 & 0 & 1 & 0 \\
 -\frac{h}{2} & 0 & 0 & -\frac{h}{2} & 0 & 0 & \frac{h}{2} & 0 & 0 & \frac{h}{2} & 0 & 0 \\
 0 & 0 & 1 & 0 & 0 & 1 & 0 & 0 & 1 & 0 & 0 & 1 \\
\end{array}
\right]^{\textrm{T}}
\end{equation}
The unit cell (see Fig.~4) consists of four beam elements. The faces are modeled by normal Euler--Bernoulli beam elements which are connected to the nodes $1\rightarrow 2$ and $4\rightarrow 3$ (i.e. $\textrm{1st node}\rightarrow\textrm{2nd node}$). The webs are modeled by Euler--Bernoulli beam elements which have rotational springs at their ends \cite{chen2005} and are connected to the nodes $2\rightarrow 3$, $1\rightarrow4$.
\section{General solution to equilibrium equations}
When the constitutive matrix (28) is used in the constitutive equations (21), the equilibrium equations (5) read
\begin{align}
&C_{11}u_x''+C_{12}\phi''+C_{15}\psi''=0, \\
&C_{12}u_x''+C_{22}\phi''+C_{25}\psi''-C_{33}(u_y'+\phi)+2C_{34}(\phi+\psi)+C_{44}(u_y'-\phi-2\psi)=0, \\
&C_{33}(u_y''+\phi')+2C_{34}(u_y''-\psi')+C_{44}(u_y''-\phi'-2\psi')=-q, \\
&C_{15}u_x''+C_{25}\phi''+C_{55}\psi''+2C_{34}(u_y'+\phi)+2C_{44}(u_y'-\phi-2\psi)=-m.
\end{align}
In addition, we have
\begin{equation}
(C_{12}-C_{15})u_x'''+(C_{22}-C_{25})\phi'''+(C_{25}-C_{55})\psi'''=m'-q
\end{equation}
from
\begin{equation}
M_x''-P_{xz}''=m'-q.
\end{equation}
The general solution procedure that leads to Eqs.~(29)--(32) contains lengthy explicit expressions and is not presented here in full. It is convenient to use mathematical software such as Maple or Mathematica to derive the solution. The relevant steps of the solution can be summarized as follows:
\begin{enumerate}
    \item Solve (B.1), (B.3) and (B.5) for $u_x''$, $u_y''$ and $\psi'''$, respectively.
    \item Differentiate (B.4) three times with respect to $x$ and then in combination with the previous step obtain an uncoupled fifth-order differential equation for $\phi$ that yields Eq.~(31).
    \item Differentiate (B.4) once with respect to $x$ to obtain an uncoupled first-order differential equation for $\psi$. The solution for $\psi$ includes a constant, say, $D_1$ to be solved later.
    \item Substitute $\phi$ and $\psi$ into (B.1) and solve the equation for $u_x$. The result is Eq.~(29).
    \item Solve (B.2) for $u_y'$ and substitute the result into (B.4) and then solve $D_1$. Final form of Eq.~(32) is found.
    \item Integrate $u_y'$ from previous step to obtain Eq.~(30). The resulting integration constant is $c_3$.
\end{enumerate}
The constant coefficients $\alpha_j$ and $\beta_j$ ($j=1,2,3$) in the solution (29)--(32) are
\begin{align}
\alpha_1&=\frac{C_{12} (C_{55}-C_{25})+C_{15} (C_{22}-C_{25})}{C_{11} (C_{25}-C_{55})+C_{15}
   (C_{15}-C_{12})} , \\
\alpha_2&=\frac{C_{11} (C_{22}-C_{25})+C_{12} (C_{15}-C_{12})}{C_{11} (C_{55}-C_{25})+C_{15}
   (C_{12}-C_{15})} , \\
\alpha_3&=\frac{\zeta_1-\zeta_2}{2 C_{11} \left(C_{34}^2-C_{33} C_{44}\right)} ,
\end{align}
where
\begin{align}
\zeta_1&=(C_{12}-C_{15})\left[C_{15}(C_{33}-C_{44})+2C_{12}(C_{34}+C_{44})\right] , \\
\zeta_2&=C_{11}\left[C_{25}(C_{33}-2C_{34}-3C_{44})+2C_{22}(C_{34}+C_{44})+C_{55}(C_{44}-C_{33})\right] .
\end{align}
and
\begin{align}
\beta_1&=\frac{2 \sqrt{C_{33} C_{44}-C_{34}^2} \sqrt{C_{11} (C_{22}-2
   C_{25}+C_{55})-(C_{12}-C_{15})^2}}{\sqrt{C_{33}+2 C_{34}+C_{44}} \sqrt{C_{11} \left(C_{22}
   C_{55}-C_{25}^2\right)+C_{12} (2 C_{15} C_{25}-C_{12} C_{55})-C_{15}^2 C_{22}}}, \\
\beta_2&=\frac{C_{33}-C_{44}-2(C_{34}+C_{44})\alpha_2}{C_{33}+2 C_{34}+C_{44}}, \\
\beta_3&=\frac{\zeta_3-\zeta_4}{C_{11} \left(C_{34}^2-C_{33} C_{44}\right)},
\end{align}
where
\begin{align}
\zeta_3&=(C_{12}-C_{15})\left[2C_{12}C_{44}+C_{15}(C_{34}-C_{44})\right] , \\
\zeta_4&=C_{11}\left[2C_{22}C_{44}+C_{25}(C_{34}-3C_{44})+C_{55}(C_{44}-C_{34})\right].
\end{align}
The particular solutions to be added to the homogeneous solution in the case of uniformly distributed loads $q(x)=q_0$ and $m(x)=m_0$ are
\begin{align}
q_{ux}&=\frac{q_0(C_{15}-C_{12})x^3}{\zeta_5}, \quad q_{uy}=\frac{q_0C_{11}(6\beta_3-x^2)x^2}{4\zeta_5} , \\
q_\phi&=\frac{q_0C_{11}x^3}{\zeta_5}, \quad q_\psi=\frac{q_0C_{11}(3\alpha_3 x-x^3)}{\zeta_5}
\end{align}
and
\begin{equation}
m_{uy}=\frac{m_0(C_{44}-C_{34})x}{2(C_{34}^2-C_{33}C_{44})}, \quad m_\psi=\frac{m_0(C_{44}-C_{34})}{4(C_{34}^2-C_{33}C_{44})}, \quad m_{ux}=m_{\phi}=0,
\end{equation}
where
\begin{equation}
\zeta_5=6[(C_{12}-C_{15})^2-C_{11}(C_{22}-2C_{25}+C_{55})].
\end{equation}
For example, $q_{ux}$ and $q_{uy}$ are added to the right-hand side of Eqs. (29) and (30), respectively.

\section*{References}
\bibliographystyle{model3-num-names}
\bibliography{microlitera}

\begin{thebibliography}{52}
\providecommand{\natexlab}[1]{#1}
\providecommand{\url}[1]{\texttt{#1}}
\providecommand{\urlprefix}{URL }
\expandafter\ifx\csname urlstyle\endcsname\relax
  \providecommand{\doi}[1]{doi:\discretionary{}{}{}#1}\else
  \providecommand{\doi}{doi:\discretionary{}{}{}\begingroup
  \urlstyle{rm}\Url}\fi
\providecommand{\eprint}[2][]{\url{#2}}
\providecommand{\BIBand}{and}
\providecommand{\bibinfo}[2]{#2}
\ifx\xfnm\undefined \def\xfnm[#1]{\unskip,\space#1}\fi
\bibitem[{Roland and Metschkow(1997)}]{roland1997}
\bibinfo{author}{Roland\xfnm[ F.]}, \bibinfo{author}{Metschkow\xfnm[ B.]}.
\newblock \bibinfo{title}{Laser welded sandwich panels for shipbuilding and
  structural steel engineering}.
\newblock \bibinfo{publisher}{Transactions on the Built Environment, vol 24.
  WIT Press}; \bibinfo{year}{1997}.
\bibitem[{Kujala and Klanac(2005)}]{kujala2005}
\bibinfo{author}{Kujala\xfnm[ P.]}, \bibinfo{author}{Klanac\xfnm[ A.]}.
\newblock \bibinfo{title}{Steel sandwich panels in marine applications}.
\newblock \bibinfo{journal}{Brodogradnja}
  \bibinfo{year}{2005};\bibinfo{volume}{56}(\bibinfo{number}{4}):\bibinfo{pages}{305--314}.
\bibitem[{Kolsters and Zenkert(2010)}]{kolsters2010}
\bibinfo{author}{Kolsters\xfnm[ H.]}, \bibinfo{author}{Zenkert\xfnm[ D.]}.
\newblock \bibinfo{title}{Buckling of laser-welded sandwich panels: ultimate
  strength and experiments}.
\newblock \bibinfo{journal}{P I Mech Eng M-J Eng}
  \bibinfo{year}{2010};\bibinfo{volume}{224}(\bibinfo{number}{1}):\bibinfo{pages}{29--45}.
\bibitem[{Jelovica et~al.(2012)Jelovica, Romanoff, Ehlers and
  Varsta}]{jelovica2012}
\bibinfo{author}{Jelovica\xfnm[ J.]}, \bibinfo{author}{Romanoff\xfnm[ J.]},
  \bibinfo{author}{Ehlers\xfnm[ S.]}, \bibinfo{author}{Varsta\xfnm[ P.]}.
\newblock \bibinfo{title}{Influence of weld stiffness on buckling strength of
  laser-welded web-core sandwich plates}.
\newblock \bibinfo{journal}{J Constr Steel R}
  \bibinfo{year}{2012};\bibinfo{volume}{77}:\bibinfo{pages}{12--18}.
\bibitem[{Jelovica et~al.(2013)Jelovica, Romanoff, Ehlers and
  Aromaa}]{jelovica2013}
\bibinfo{author}{Jelovica\xfnm[ J.]}, \bibinfo{author}{Romanoff\xfnm[ J.]},
  \bibinfo{author}{Ehlers\xfnm[ S.]}, \bibinfo{author}{Aromaa\xfnm[ J.]}.
\newblock \bibinfo{title}{Ultimate strength of corroded web-core sandwich
  beams}.
\newblock \bibinfo{journal}{Mar Struct}
  \bibinfo{year}{2013};\bibinfo{volume}{31}:\bibinfo{pages}{1--14}.
\bibitem[{Jiang et~al.(2014)Jiang, Zhu, Qiao, Wu, Li and Chen}]{jiang2014}
\bibinfo{author}{Jiang\xfnm[ X.X.]}, \bibinfo{author}{Zhu\xfnm[ L.]},
  \bibinfo{author}{Qiao\xfnm[ J.S.]}, \bibinfo{author}{Wu\xfnm[ Y.X.]},
  \bibinfo{author}{Li\xfnm[ Z.G.]}, \bibinfo{author}{Chen\xfnm[ J.H.]}.
\newblock \bibinfo{title}{The strength of laser welded web-core steel sandwich
  plates}.
\newblock \bibinfo{journal}{Appl Mech Mater}
  \bibinfo{year}{2014};\bibinfo{volume}{551}:\bibinfo{pages}{42--46}.
\bibitem[{Jelovica et~al.(2014)Jelovica, Romanoff and Remes}]{jelovica2014}
\bibinfo{author}{Jelovica\xfnm[ J.]}, \bibinfo{author}{Romanoff\xfnm[ J.]},
  \bibinfo{author}{Remes\xfnm[ H.]}.
\newblock \bibinfo{title}{Influence of general corrosion on buckling strength
  of laser-welded web-core sandwich plates}.
\newblock \bibinfo{journal}{J Constr Steel R}
  \bibinfo{year}{2014};\bibinfo{volume}{101}:\bibinfo{pages}{342--350}.
\bibitem[{Frank et~al.(2013)Frank, Romanoff and Remes}]{frank2013}
\bibinfo{author}{Frank\xfnm[ D.]}, \bibinfo{author}{Romanoff\xfnm[ J.]},
  \bibinfo{author}{Remes\xfnm[ H.]}.
\newblock \bibinfo{title}{Fatigue strength assessment of laser stake-welded
  web-core steel sandwich panels}.
\newblock \bibinfo{journal}{Fatigue Fract Eng M}
  \bibinfo{year}{2013};\bibinfo{volume}{36}(\bibinfo{number}{8}):\bibinfo{pages}{724--737}.
\bibitem[{Remes et~al.(2017)Remes, Romanoff, Lillem{\"a}e, Frank, Liinalampi,
  Lehto et~al.}]{remes2017}
\bibinfo{author}{Remes\xfnm[ H.]}, \bibinfo{author}{Romanoff\xfnm[ J.]},
  \bibinfo{author}{Lillem{\"a}e\xfnm[ I.]}, \bibinfo{author}{Frank\xfnm[ D.]},
  \bibinfo{author}{Liinalampi\xfnm[ S.]}, \bibinfo{author}{Lehto\xfnm[ P.]},
  \bibinfo{author}{Varsta\xfnm[ P.]}.
\newblock \bibinfo{title}{Factors affecting the fatigue strength of thin-plates
  in large structures}.
\newblock \bibinfo{journal}{Int J Fatigue}
  \bibinfo{year}{2017};\bibinfo{volume}{101}:\bibinfo{pages}{397--407}.
\bibitem[{Gallo et~al.(2018)Gallo, Guglielmo, Romanoff and Remes}]{gallo2018}
\bibinfo{author}{Gallo\xfnm[ P.]}, \bibinfo{author}{Guglielmo\xfnm[ M.]},
  \bibinfo{author}{Romanoff\xfnm[ J.]}, \bibinfo{author}{Remes\xfnm[ H.]}.
\newblock \bibinfo{title}{Influence of crack tip plasticity on fatigue
  behaviour of laser stake-welded {T}-joints made of thin plates}.
\newblock \bibinfo{journal}{Int J Mech Sci}
  \bibinfo{year}{2018};\bibinfo{volume}{136}:\bibinfo{pages}{112--123}.
\bibitem[{Bright and Smith(2004)}]{bright2004}
\bibinfo{author}{Bright\xfnm[ S.R.]}, \bibinfo{author}{Smith\xfnm[ J.W.]}.
\newblock \bibinfo{title}{Fatigue performance of laser-welded steel bridge
  decks}.
\newblock \bibinfo{journal}{Struct Eng}
  \bibinfo{year}{2004};\bibinfo{volume}{82}(\bibinfo{number}{21}).
\bibitem[{Bright and Smith(2007)}]{bright2007}
\bibinfo{author}{Bright\xfnm[ S.R.]}, \bibinfo{author}{Smith\xfnm[ J.W.]}.
\newblock \bibinfo{title}{A new design for steel bridge decks using laser
  fabrication}.
\newblock \bibinfo{journal}{Struct Eng}
  \bibinfo{year}{2007};\bibinfo{volume}{85}(\bibinfo{number}{21}).
\bibitem[{Nilsson et~al.(2017)Nilsson, Al-Emrani and Atashipour}]{nilsson2017}
\bibinfo{author}{Nilsson\xfnm[ P.]}, \bibinfo{author}{Al-Emrani\xfnm[ M.]},
  \bibinfo{author}{Atashipour\xfnm[ S.R.]}.
\newblock \bibinfo{title}{Transverse shear stiffness of corrugated core steel
  sandwich panels with dual weld lines}.
\newblock \bibinfo{journal}{Thin Wall Struct}
  \bibinfo{year}{2017};\bibinfo{volume}{117}:\bibinfo{pages}{98--112}.
\bibitem[{Briscoe et~al.(2011)Briscoe, Mantell, Davidson and
  Okazaki}]{briscoe2011}
\bibinfo{author}{Briscoe\xfnm[ C.R.]}, \bibinfo{author}{Mantell\xfnm[ S.C.]},
  \bibinfo{author}{Davidson\xfnm[ J.H.]}, \bibinfo{author}{Okazaki\xfnm[ T.]}.
\newblock \bibinfo{title}{Design procedure for web core sandwich panels for
  residential roofs}.
\newblock \bibinfo{journal}{J Sandw Struct Mater}
  \bibinfo{year}{2011};\bibinfo{volume}{13}(\bibinfo{number}{1}):\bibinfo{pages}{23--58}.
\bibitem[{Reddy(2004)}]{reddy2004}
\bibinfo{author}{Reddy\xfnm[ J.N.]}.
\newblock \bibinfo{title}{Mechanics of Laminated Composite Plates and Shells:
  Theory and Analysis}.
\newblock \bibinfo{year}{2004}.
\bibitem[{Romanoff et~al.(2007{\natexlab{a}})Romanoff, Varsta and
  Klanac}]{romanoff2007a}
\bibinfo{author}{Romanoff\xfnm[ J.]}, \bibinfo{author}{Varsta\xfnm[ P.]},
  \bibinfo{author}{Klanac\xfnm[ A.]}.
\newblock \bibinfo{title}{Stress analysis of homogenized web-core sandwich
  beams}.
\newblock \bibinfo{journal}{Compos Struct}
  \bibinfo{year}{2007}{\natexlab{a}};\bibinfo{volume}{79}(\bibinfo{number}{3}):\bibinfo{pages}{411--422}.
\bibitem[{Romanoff and Varsta(2007)}]{romanoff2007b}
\bibinfo{author}{Romanoff\xfnm[ J.]}, \bibinfo{author}{Varsta\xfnm[ P.]}.
\newblock \bibinfo{title}{Bending response of web-core sandwich plates}.
\newblock \bibinfo{journal}{Compos Struct}
  \bibinfo{year}{2007};\bibinfo{volume}{81}(\bibinfo{number}{2}):\bibinfo{pages}{292--302}.
\bibitem[{Romanoff and Reddy(2014)}]{romanoff2014}
\bibinfo{author}{Romanoff\xfnm[ J.]}, \bibinfo{author}{Reddy\xfnm[ J.N.]}.
\newblock \bibinfo{title}{Experimental validation of the modified couple stress
  {T}imoshenko beam theory for web-core sandwich panels}.
\newblock \bibinfo{journal}{Compos Struct}
  \bibinfo{year}{2014};\bibinfo{volume}{111}:\bibinfo{pages}{130--137}.
\bibitem[{Romanoff et~al.(2016)Romanoff, Reddy and Jelovica}]{romanoff2016}
\bibinfo{author}{Romanoff\xfnm[ J.]}, \bibinfo{author}{Reddy\xfnm[ J.N.]},
  \bibinfo{author}{Jelovica\xfnm[ J.]}.
\newblock \bibinfo{title}{Using non-local {T}imoshenko beam theories for
  prediction of micro-and macro-structural responses}.
\newblock \bibinfo{journal}{Compos Struct}
  \bibinfo{year}{2016};\bibinfo{volume}{156}:\bibinfo{pages}{410--420}.
\bibitem[{Gesualdo et~al.(2017)Gesualdo, Iannuzzo, Penta and
  Pucillo}]{gesualdo2017}
\bibinfo{author}{Gesualdo\xfnm[ A.]}, \bibinfo{author}{Iannuzzo\xfnm[ A.]},
  \bibinfo{author}{Penta\xfnm[ F.]}, \bibinfo{author}{Pucillo\xfnm[ G.P.]}.
\newblock \bibinfo{title}{Homogenization of a {V}ierendeel girder with elastic
  joints into an equivalent polar beam}.
\newblock \bibinfo{journal}{J Mech Mater Struct}
  \bibinfo{year}{2017};\bibinfo{volume}{12}(\bibinfo{number}{4}):\bibinfo{pages}{485--504}.
\bibitem[{Penta et~al.(2017)Penta, Monaco, Pucillo and Gesualdo}]{penta2017}
\bibinfo{author}{Penta\xfnm[ F.]}, \bibinfo{author}{Monaco\xfnm[ M.]},
  \bibinfo{author}{Pucillo\xfnm[ G.P.]}, \bibinfo{author}{Gesualdo\xfnm[ A.]}.
\newblock \bibinfo{title}{Periodic beam-like structures homogenization by
  transfer matrix eigen-analysis: A direct approach}.
\newblock \bibinfo{journal}{Mech Res Commun}
  \bibinfo{year}{2017};\bibinfo{volume}{85}:\bibinfo{pages}{81--88}.
\bibitem[{Karttunen et~al.(2018)Karttunen, Reddy and Romanoff}]{karttunen2018a}
\bibinfo{author}{Karttunen\xfnm[ A.T.]}, \bibinfo{author}{Reddy\xfnm[ J.N.]},
  \bibinfo{author}{Romanoff\xfnm[ J.]}.
\newblock \bibinfo{title}{Micropolar modeling approach for periodic sandwich
  beams}.
\newblock \bibinfo{journal}{Compos Struct}
  \bibinfo{year}{2018};\bibinfo{volume}{185}:\bibinfo{pages}{656--664}.
\bibitem[{Pydah and Bhaskar(2016)}]{pydah2016}
\bibinfo{author}{Pydah\xfnm[ A.]}, \bibinfo{author}{Bhaskar\xfnm[ K.]}.
\newblock \bibinfo{title}{An accurate discrete model for web-core sandwich
  plates}.
\newblock \bibinfo{journal}{J Sandw Struct Mater}
  \bibinfo{year}{2016};\bibinfo{volume}{18}(\bibinfo{number}{4}):\bibinfo{pages}{474--500}.
\bibitem[{Pydah and Bhaskar(2017)}]{pydah2017}
\bibinfo{author}{Pydah\xfnm[ A.]}, \bibinfo{author}{Bhaskar\xfnm[ K.]}.
\newblock \bibinfo{title}{Accurate analytical solutions for shear-deformable
  web-core sandwich plates}.
\newblock \bibinfo{journal}{J Sandw Struct Mater}
  \bibinfo{year}{2017};\bibinfo{volume}{19}(\bibinfo{number}{5}):\bibinfo{pages}{616--643}.
\bibitem[{Pydah and Batra(2018)}]{pydah2018}
\bibinfo{author}{Pydah\xfnm[ A.]}, \bibinfo{author}{Batra\xfnm[ R.]}.
\newblock \bibinfo{title}{Analytical solution for cylindrical bending of
  two-layered corrugated and webcore sandwich panels}.
\newblock \bibinfo{journal}{Thin Wall Struct}
  \bibinfo{year}{2018};\bibinfo{volume}{123}:\bibinfo{pages}{509--519}.
\bibitem[{Noor and Nemeth(1980)}]{noor1980}
\bibinfo{author}{Noor\xfnm[ A.K.]}, \bibinfo{author}{Nemeth\xfnm[ M.P.]}.
\newblock \bibinfo{title}{Micropolar beam models for lattice grids with rigid
  joints}.
\newblock \bibinfo{journal}{Comput Meth Appl Mech Eng}
  \bibinfo{year}{1980};\bibinfo{volume}{21}(\bibinfo{number}{2}):\bibinfo{pages}{249--263}.
\bibitem[{Noor(1988)}]{noor1988}
\bibinfo{author}{Noor\xfnm[ A.K.]}.
\newblock \bibinfo{title}{Continuum modeling for repetitive lattice
  structures}.
\newblock \bibinfo{journal}{Appl Mech Rev}
  \bibinfo{year}{1988};\bibinfo{volume}{41}(\bibinfo{number}{7}):\bibinfo{pages}{285--296}.
\bibitem[{Ostoja-Starzewski(2002)}]{ostoja2002}
\bibinfo{author}{Ostoja-Starzewski\xfnm[ M.]}.
\newblock \bibinfo{title}{Lattice models in micromechanics}.
\newblock \bibinfo{journal}{Appl Mech Rev}
  \bibinfo{year}{2002};\bibinfo{volume}{55}(\bibinfo{number}{1}):\bibinfo{pages}{35--60}.
\bibitem[{Kouznetsova et~al.(2002)Kouznetsova, Geers and
  Brekelmans}]{kouznetsova2002}
\bibinfo{author}{Kouznetsova\xfnm[ V.]}, \bibinfo{author}{Geers\xfnm[ M.G.D.]},
  \bibinfo{author}{Brekelmans\xfnm[ W.A.M.]}.
\newblock \bibinfo{title}{Multi-scale constitutive modelling of heterogeneous
  materials with a gradient-enhanced computational homogenization scheme}.
\newblock \bibinfo{journal}{Int J Numer Meth Eng}
  \bibinfo{year}{2002};\bibinfo{volume}{54}(\bibinfo{number}{8}):\bibinfo{pages}{1235--1260}.
\bibitem[{Larsson and Diebels(2007)}]{larsson2007}
\bibinfo{author}{Larsson\xfnm[ R.]}, \bibinfo{author}{Diebels\xfnm[ S.]}.
\newblock \bibinfo{title}{A second-order homogenization procedure for
  multi-scale analysis based on micropolar kinematics}.
\newblock \bibinfo{journal}{Int J Numer Meth Eng}
  \bibinfo{year}{2007};\bibinfo{volume}{69}(\bibinfo{number}{12}):\bibinfo{pages}{2485--2512}.
\bibitem[{Geers et~al.(2010)Geers, Kouznetsova and Brekelmans}]{geers2010}
\bibinfo{author}{Geers\xfnm[ M.G.D.]}, \bibinfo{author}{Kouznetsova\xfnm[
  V.G.]}, \bibinfo{author}{Brekelmans\xfnm[ W.A.M.]}.
\newblock \bibinfo{title}{Multi-scale computational homogenization: Trends and
  challenges}.
\newblock \bibinfo{journal}{J Comput Appl Math}
  \bibinfo{year}{2010};\bibinfo{volume}{234}(\bibinfo{number}{7}):\bibinfo{pages}{2175--2182}.
\bibitem[{Matou{\v{s}} et~al.(2017)Matou{\v{s}}, Geers, Kouznetsova and
  Gillman}]{matouvs2017}
\bibinfo{author}{Matou{\v{s}}\xfnm[ K.]}, \bibinfo{author}{Geers\xfnm[
  M.G.D.]}, \bibinfo{author}{Kouznetsova\xfnm[ V.G.]},
  \bibinfo{author}{Gillman\xfnm[ A.]}.
\newblock \bibinfo{title}{A review of predictive nonlinear theories for
  multiscale modeling of heterogeneous materials}.
\newblock \bibinfo{journal}{J Comput Phys}
  \bibinfo{year}{2017};\bibinfo{volume}{330}:\bibinfo{pages}{192--220}.
\bibitem[{Kumar and McDowell(2004)}]{kumar2004}
\bibinfo{author}{Kumar\xfnm[ R.S.]}, \bibinfo{author}{McDowell\xfnm[ D.L.]}.
\newblock \bibinfo{title}{Generalized continuum modeling of 2-{D} periodic
  cellular solids}.
\newblock \bibinfo{journal}{Int J Solids Struct}
  \bibinfo{year}{2004};\bibinfo{volume}{41}(\bibinfo{number}{26}):\bibinfo{pages}{7399--7422}.
\bibitem[{Spadoni and Ruzzene(2012)}]{spadoni2012}
\bibinfo{author}{Spadoni\xfnm[ A.]}, \bibinfo{author}{Ruzzene\xfnm[ M.]}.
\newblock \bibinfo{title}{Elasto-static micropolar behavior of a chiral auxetic
  lattice}.
\newblock \bibinfo{journal}{J Mech Phys Solids}
  \bibinfo{year}{2012};\bibinfo{volume}{60}(\bibinfo{number}{1}):\bibinfo{pages}{156--171}.
\bibitem[{Trovalusci et~al.(2017)Trovalusci, De~Bellis and
  Masiani}]{trovalusci2017}
\bibinfo{author}{Trovalusci\xfnm[ P.]}, \bibinfo{author}{De~Bellis M.\xfnm[
  L.]}, \bibinfo{author}{Masiani\xfnm[ R.]}.
\newblock \bibinfo{title}{A multiscale description of particle composites: From
  lattice microstructures to micropolar continua}.
\newblock \bibinfo{journal}{Compos B-Eng}
  \bibinfo{year}{2017};\bibinfo{volume}{128}:\bibinfo{pages}{164--173}.
\bibitem[{Barber(2010)}]{barber2010}
\bibinfo{author}{Barber\xfnm[ J.R.]}.
\newblock \bibinfo{title}{{Elasticity}}.
\newblock \bibinfo{address}{New York}: \bibinfo{publisher}{Springer};
  \bibinfo{edition}{3rd} ed.; \bibinfo{year}{2010}.
\bibitem[{Monforton and Wu(1963)}]{monforton1963}
\bibinfo{author}{Monforton\xfnm[ G.R.]}, \bibinfo{author}{Wu\xfnm[ T.H.]}.
\newblock \bibinfo{title}{Matrix analysis of semi-rigid connected frames}.
\newblock \bibinfo{journal}{J Struct Div-ASCE}
  \bibinfo{year}{1963};\bibinfo{volume}{89}(\bibinfo{number}{6}):\bibinfo{pages}{13--24}.
\bibitem[{Chen and Lui(2005)}]{chen2005}
\bibinfo{author}{Chen\xfnm[ W.H.]}, \bibinfo{author}{Lui\xfnm[ E.M.]}.
\newblock \bibinfo{title}{Handbook of structural engineering}.
\newblock \bibinfo{publisher}{CRC Press}; \bibinfo{year}{2005}.
\bibitem[{Romanoff et~al.(2007{\natexlab{b}})Romanoff, Remes, Socha, Jutila and
  Varsta}]{romanoff2007c}
\bibinfo{author}{Romanoff\xfnm[ J.]}, \bibinfo{author}{Remes\xfnm[ H.]},
  \bibinfo{author}{Socha\xfnm[ G.]}, \bibinfo{author}{Jutila\xfnm[ M.]},
  \bibinfo{author}{Varsta\xfnm[ P.]}.
\newblock \bibinfo{title}{The stiffness of laser stake welded {T}-joints in
  web-core sandwich structures}.
\newblock \bibinfo{journal}{Thin Wall Struct}
  \bibinfo{year}{2007}{\natexlab{b}};\bibinfo{volume}{45}(\bibinfo{number}{4}):\bibinfo{pages}{453--462}.
\bibitem[{Ting(1996)}]{ting1996}
\bibinfo{author}{Ting\xfnm[ T.C.T.]}.
\newblock \bibinfo{title}{Anisotropic elasticity: theory and applications}.
\newblock \bibinfo{number}{45}; \bibinfo{publisher}{Oxford University Press};
  \bibinfo{year}{1996}.
\bibitem[{Eringen(2012)}]{eringen2012}
\bibinfo{author}{Eringen\xfnm[ A.C.]}.
\newblock \bibinfo{title}{Microcontinuum Field Theories: I. Foundations and
  Solids}.
\newblock \bibinfo{publisher}{Springer Science \& Business Media};
  \bibinfo{year}{2012}.
\bibitem[{Allen(1969)}]{allen1969}
\bibinfo{author}{Allen\xfnm[ H.G.]}.
\newblock \bibinfo{title}{Analysis and Design of Structural Sandwich Panels}.
\newblock \bibinfo{publisher}{Pergamon Press}; \bibinfo{year}{1969}.
\bibitem[{Karttunen and von Hertzen(2016)}]{karttunen2016c}
\bibinfo{author}{Karttunen\xfnm[ A.T.]}, \bibinfo{author}{von Hertzen\xfnm[
  R.]}.
\newblock \bibinfo{title}{On the foundations of anisotropic interior beam
  theories}.
\newblock \bibinfo{journal}{Compos B-Eng}
  \bibinfo{year}{2016};\bibinfo{volume}{87}:\bibinfo{pages}{299--310}.
\bibitem[{Romanoff(2014)}]{romanoff2014b}
\bibinfo{author}{Romanoff\xfnm[ J.]}.
\newblock \bibinfo{title}{Optimization of web-core steel sandwich decks at
  concept design stage using envelope surface for stress assessment}.
\newblock \bibinfo{journal}{Eng Struct}
  \bibinfo{year}{2014};\bibinfo{volume}{66}:\bibinfo{pages}{1--9}.
\bibitem[{Karttunen et~al.(2017)Karttunen, Kanerva, Frank, Romanoff, Remes,
  Jelovica et~al.}]{karttunen2017b}
\bibinfo{author}{Karttunen\xfnm[ A.T.]}, \bibinfo{author}{Kanerva\xfnm[ M.]},
  \bibinfo{author}{Frank\xfnm[ D.]}, \bibinfo{author}{Romanoff\xfnm[ J.]},
  \bibinfo{author}{Remes\xfnm[ H.]}, \bibinfo{author}{Jelovica\xfnm[ J.]},
  \bibinfo{author}{Bossuyt\xfnm[ S.]}, \bibinfo{author}{Sarlin\xfnm[ E.]}.
\newblock \bibinfo{title}{Fatigue strength of laser-welded foam-filled steel
  sandwich beams}.
\newblock \bibinfo{journal}{Mater Design}
  \bibinfo{year}{2017};\bibinfo{volume}{115}:\bibinfo{pages}{64--72}.
\bibitem[{Fleck et~al.(2010)Fleck, Deshpande and Ashby}]{fleck2010}
\bibinfo{author}{Fleck\xfnm[ N.A.]}, \bibinfo{author}{Deshpande\xfnm[ V.S.]},
  \bibinfo{author}{Ashby\xfnm[ M.F.]}.
\newblock \bibinfo{title}{Micro-architectured materials: past, present and
  future}.
\newblock \bibinfo{journal}{P Roy Soc A}
  \bibinfo{year}{2010};\bibinfo{volume}{466}:\bibinfo{pages}{2495--2516}.
\bibitem[{Goncalves et~al.(2017)Goncalves, Karttunen, Romanoff and
  Reddy}]{goncalves2017}
\bibinfo{author}{Goncalves\xfnm[ B.R.]}, \bibinfo{author}{Karttunen\xfnm[
  A.T.]}, \bibinfo{author}{Romanoff\xfnm[ J.]}, \bibinfo{author}{Reddy\xfnm[
  J.N.]}.
\newblock \bibinfo{title}{Buckling and free vibration of shear-flexible
  sandwich beams using a couple-stress-based finite element}.
\newblock \bibinfo{journal}{Compos Struct}
  \bibinfo{year}{2017};\bibinfo{volume}{165}:\bibinfo{pages}{233--241}.
\bibitem[{Jelovica et~al.(2016)Jelovica, Romanoff and Klein}]{jelovica2016}
\bibinfo{author}{Jelovica\xfnm[ J.]}, \bibinfo{author}{Romanoff\xfnm[ J.]},
  \bibinfo{author}{Klein\xfnm[ R.]}.
\newblock \bibinfo{title}{Eigenfrequency analyses of laser-welded web-core
  sandwich panels}.
\newblock \bibinfo{journal}{Thin Wall Struct}
  \bibinfo{year}{2016};\bibinfo{volume}{101}:\bibinfo{pages}{120--128}.
\bibitem[{Reddy(2007)}]{reddy2007}
\bibinfo{author}{Reddy\xfnm[ J.N.]}.
\newblock \bibinfo{title}{Nonlocal theories for bending, buckling and vibration
  of beams}.
\newblock \bibinfo{journal}{Int J Eng Sci}
  \bibinfo{year}{2007};\bibinfo{volume}{45}(\bibinfo{number}{2-8}):\bibinfo{pages}{288--307}.
\bibitem[{Ma et~al.(2008)Ma, Gao and Reddy}]{ma2008}
\bibinfo{author}{Ma\xfnm[ H.M.]}, \bibinfo{author}{Gao\xfnm[ X.L.]},
  \bibinfo{author}{Reddy\xfnm[ J.N.]}.
\newblock \bibinfo{title}{A microstructure-dependent {T}imoshenko beam model
  based on a modified couple stress theory}.
\newblock \bibinfo{journal}{J Mech Phys Solids}
  \bibinfo{year}{2008};\bibinfo{volume}{56}(\bibinfo{number}{12}):\bibinfo{pages}{3379--3391}.
\bibitem[{Reddy(2011)}]{reddy2011}
\bibinfo{author}{Reddy\xfnm[ J.N.]}.
\newblock \bibinfo{title}{Microstructure-dependent couple stress theories of
  functionally graded beams}.
\newblock \bibinfo{journal}{J Mech Phys Solids}
  \bibinfo{year}{2011};\bibinfo{volume}{59}(\bibinfo{number}{11}):\bibinfo{pages}{2382--2399}.
\bibitem[{Asghari et~al.(2011)Asghari, Rahaeifard, Kahrobaiyan and
  Ahmadian}]{asghari2011}
\bibinfo{author}{Asghari\xfnm[ M.]}, \bibinfo{author}{Rahaeifard\xfnm[ M.]},
  \bibinfo{author}{Kahrobaiyan\xfnm[ M.H.]}, \bibinfo{author}{Ahmadian\xfnm[
  M.T.]}.
\newblock \bibinfo{title}{The modified couple stress functionally graded
  {T}imoshenko beam formulation}.
\newblock \bibinfo{journal}{Mater Design}
  \bibinfo{year}{2011};\bibinfo{volume}{32}(\bibinfo{number}{3}):\bibinfo{pages}{1435--1443}.

\end{thebibliography}






\end{document}